\documentclass[prd,twocolumn,preprintnumbers,showpacs,nofootinbib,superscriptaddress,notitlepage,floatfix,longbibliography]{revtex4-1}
\usepackage{hyperref}
\usepackage{graphicx}
\usepackage{amsmath}
\usepackage{amssymb}
\usepackage{slashed}
\usepackage{amsfonts}
\usepackage{xcolor}
\usepackage{multirow}
\usepackage[caption=false]{subfig}
\usepackage{array}
\usepackage{mwe}
\usepackage{dsfont}

\usepackage[utf8]{inputenc}
\usepackage[normalem]{ulem}

\newcommand{\be}{\begin{equation}}
\newcommand{\ee}{\end{equation}}
\newcommand{\bea}{\begin{eqnarray}}
\newcommand{\eea}{\end{eqnarray}}
\newcommand{\ba}{\begin{eqnarray}}
\newcommand{\ea}{\end{eqnarray}}

\begin{document}

\title{ Photo-production of $\eta_{c,b}$ near Threshold}

\author{Wei-Yang Liu}
\email{wei-yang.liu@stonybrook.edu}
\author{Ismail Zahed}
\email{ismail.zahed@stonybrook.edu}
\affiliation{Center for Nuclear Theory, Department of Physics and Astronomy, Stony Brook University, Stony Brook, New York 11794–3800, USA}

\date{\today}

\begin{abstract}
We analyze the photoproduction of $\eta_{c,b}$ off a proton in the threshold region,
in terms of C-odd gluonic correlations in the off-forward proton matrix element.
Near threshold, the skewness is large leading to a production amplitude that is dominated by four C-odd twist-3 gluon GPDs.  We use the QCD instanton vacuum to estimate these C-odd contributions  in the proton. The results are used to estimate the differential cross sections for coherent photo-production of $\eta_{c,b}$ in the threshold region, at current  electron facilities. 
\end{abstract}

\maketitle

\section{Introduction}
The origin of mass and intrinsic spin in hadronic physics, is intimately related
to the non-perturbative aspects of the QCD vacuum. At low resolution, this vacuum is populated with topologically active gluonic configurations, in the form of tunneling  instantons and anti-instantons (pseudoparticles)~\cite{Schafer:1996wv}. These pseudoparticles are classical solutions to the Yang-Mills equations, 
that interpolate between topologically distinct vacua. They break conformal symmetry and chiral symmetry, producing mass from no mass~\cite{Shuryak:2018fjr,Zahed:2021fxk} (and references therein). 
These pseudoparticles are the dominant component of the vacuum in
gluodynamics at low resolution, as revealed numerically by cooling
procedures using gradient flow~\cite{Leinweber:1999cw,Biddle:2023lod}.

 Recent experiments carried at JLAB~\cite{Hafidi:2017bsg,GlueX:2019mkq,Meziani:2020oks,Anderle:2021wcy} may have started to
reveal some aspects of this primordial glue. Near threshold diffractive photo- or electro-production of  heavy mesons such as charmonium or bottomonium, maybe exclusively sensitive
to the glue content of the nucleon as scooped from the QCD vacuum.  The recently reported data by JLAB have spurred a renewed interest by many~\cite{Hatta:2018ina,Mamo:2019mka,Kharzeev:2021qkd,Ji:2021mtz,Hatta:2021can,Guo:2021ibg,Sun:2021gmi,Wang:2022vhr}.

That gluons dominate the diffractive vector meson production at large center of mass energy $\sqrt s$ is
not surprising. What is a bit surprising, is that they may still dominate the threshold production of heavy quarkonia. Indeed, at large $\sqrt s$ diffractive $pp$  and $p\bar p$ is dominated by Pomeron exchange a tower a C-even soft gluons with positive signature, with a small Odderon admixture~\cite{Bartels:1980pe,Kwiecinski:1980wb,Braun:1998fs}, a tower of C-odd soft gluons with negative signature~\cite{TOTEM:2020zzr}.  Negative signature  Reggeons add in the $pp$ channel, and subtract in the $p\bar p$ channel.

threshold $J/\Psi$ photo-production at JLAB has allowed for a measurement of the gluonic gravitational form factors of the proton~\cite{GlueX:2019mkq,Meziani:2020oks}. More specifically,
the tensor $A$ and scalar $D$ form-factors, with the latter providing for a potential map of the gluonic shear and pressure content of the proton. The $A$-form factor is dominated by the 
$2^{++}$ glueball exchange, and the $D$-form factor is a balance between the $2^{++}$ and $0^{++}$ glueball exchanges. The Reggeized $2^{++}$ trajectory gives rise to the Pomeron, a C-even 
gluon exchange.

Threshold $\eta_c$ photoproduction on a proton, if detectable at JLAB or other facilities, probes C-odd gluon exchanges. The latters Reggeize to the
Odderon at higher center of mass energy. The aim of this paper is to analyze the near threshold production of the $\eta_{c,b}$, by factorizing out the
leading gluonic contribution to the production process, in the large skewness limit. This C-odd
and twist-3 contribution in the proton, is then evaluated in the QCD instanton vacuum. For completeness, we note a recent
analysis of the process in dual gravity~\cite{Hechenberger:2024abg}.

The outline of the paper is as follows: in section~\ref{SEC2} we review the key aspects 
of heavy charm and eta kinematics in the photo-production process on the nucleon. We generalize
the QCD arguments used for $J/\Psi$ production to $\eta_c$ production. For large skewness, the production mechanism is dominated by C-odd 3-gluon exchanges.
In section~\ref{SEC3} we detail the evaluation of the leading gluonic matrix elements in the proton, 
in the QCD instanton vacuum in leading order in the
instanton density. In section~\ref{SEC4} we discuss
the total and differential cross sections for C-even $J/\Psi, \Upsilon$ and C-odd $\eta_c,\eta_b$
near threshold production, which allow for a parallel comparison at current electron facilities.
Our conclusions are in section~\ref{SEC5}. Additional details regarding parts of the derivation are given in the Appendices.


\section{Meson Photoproduction Near Threshold}
\label{SEC2}

\subsection{Kinematics}
\begin{figure}
    \centering
    \includegraphics[scale=0.8]{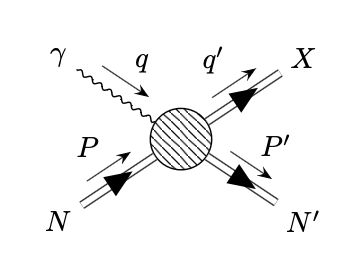}
    \caption{Kinematics for the $\gamma N\rightarrow XN'$ process.}
    \label{fig:scattering}
\end{figure}

The kinematics of the threshold production is captured by the Mandelstam invariants $s,t$, with  $s=(P+q)^2$ related to the center of mass energy $W=\sqrt{s}$, and $t=\Delta^2$  related to the momentum transfer $\Delta^\mu=(P'-P)^\mu$. The  $Q^2$ in photoproduction is exactly set to be $0$ although similar analysis can be easily extended for the large-$Q^2$ leptoproduction.
Without loss of generality, we can work in the center of mass frame. In Fig.\ref{fig:scattering}, The four-momenta of the incoming photon, incoming proton, outgoing proton and outgoing meson $X$ are denoted by $q$, $P$, $P'$, and $q'$ respectively. Each external state is given by the on-shell conditions defined as
\begin{align*}
P^2=P'^2&=M_N^2\ , & q^2=&0\ , & q'^2=M_X^2
\end{align*} 
With the on-shell conditions, the four-momenta in the center of mass frame, can be written as
\begin{align}
\label{eq:kinematics}
q=&\left(\frac{s-M^2_N}{2\sqrt{s}},\ 0,\ -\frac{s-M^2_N}{2\sqrt{s}}\right)  & \\[5pt] \nonumber 
q'=&\left(\frac{s+M_X^2-M^2_N}{2\sqrt{s}},\ -|\vec{P}'_{c}|\sin\theta,\ -|\vec{P}'_{c}|\cos\theta\right) \\[5pt] \nonumber
P=&\left(\frac{s+M^2_N}{2\sqrt{s}},\ 0,\ \frac{s-M^2_N}{2\sqrt{s}}\right)  &  \\[5pt] \nonumber
P'=&\left(\frac{s-M_X^2+M^2_N}{2\sqrt{s}},|\vec{P}'_{c}|\sin\theta,\ |\vec{P}'_{c}|\cos\theta \right)
\end{align}
where $M_N$ is the nucleon mass, $M_X$ is the produced meson mass, and $\theta$ is the scattering angle in the center of mass frame. 
The magnitude of the outgoing three-momentum reads
\begin{equation}
|\vec{P}'_c|=\left(\frac{[s-(M_X+M_N)^2][s-(M_X-M_N)^2]}{4s}\right)^{1/2}    
\end{equation}   

\begin{figure*}
\subfloat[\label{fig:t-W_c}]{%
\includegraphics[height=5cm,width=.45\linewidth]{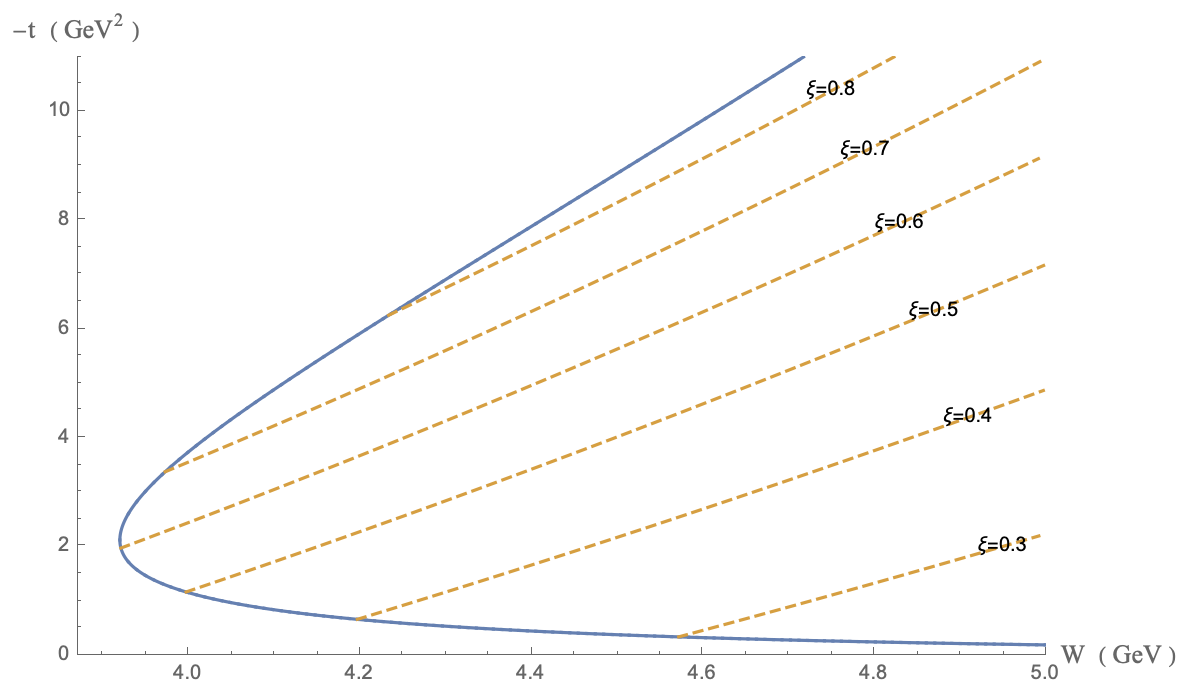}%
}\hfill
\subfloat[\label{fig:t-W_b}]{%
\includegraphics[height=5cm,width=.45\linewidth]{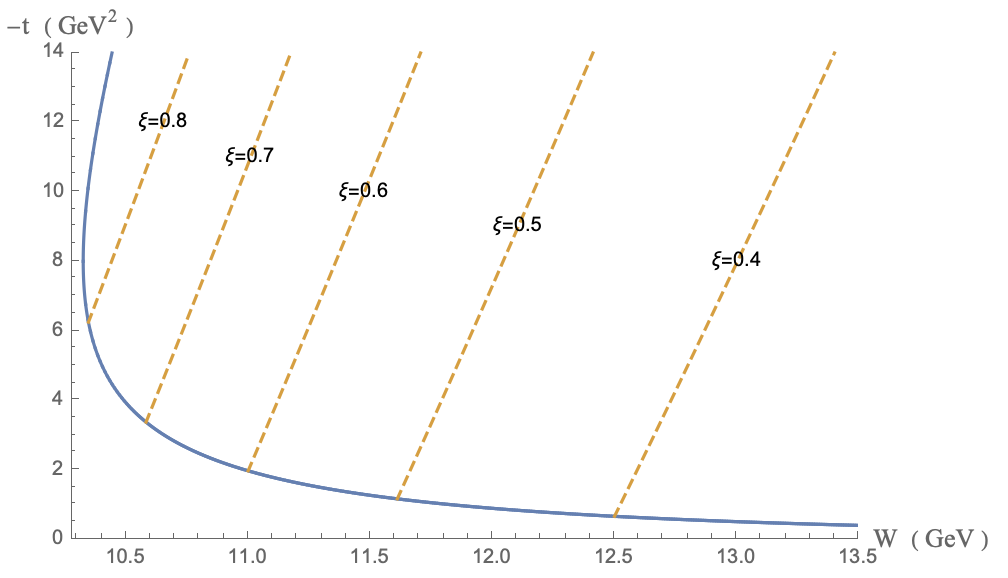}%
}
\caption{ a: Skewness $\xi$ in  the $(W,-t)$ plane in the kinematically allowed region with $M_{\eta_c}=2.982$ GeV. The kinematically allowed region for charmonium $J/\psi$ is similar due to the similar mass  $M_{J/\psi}=3.097$ GeV.;
b: Skewness $\xi$ in the $(W,-t)$ plane in the kinematically allowed region with $M_{\eta_b}=9.388$ GeV. The kinematically allowed region for bottomonium $\Upsilon$ is similar due to the similar mass $M_{\Upsilon}=9.46$ GeV.}
\end{figure*}

The scattering angle is fixed by the invariant $t$,
\begin{equation}
    \cos\theta=\frac{2st+(s-M^2_N)^2-M^2_X(s+M_N^2)}{2\sqrt{s}|\vec{P}'_c|(s-M_N^2)}
\end{equation}
Also, the skewness $\xi$ is
\begin{equation}
    \xi=-\frac{\Delta\cdot q}{2\bar{P}\cdot q}
\end{equation}
where $\bar{P}^\mu=(P+P')^\mu/2$.
In the threshold limit $\sqrt{s}\rightarrow M_N+M_X$, the momentum transfer $t$ is constrained in the neighborhood of $t_{th}$
\begin{equation}
\label{eq:t_th}
    t_{th}=-\frac{M_NM_X^2}{M_N+M_X}
\end{equation}
The kinematically allowed regions are shown on the $(W,-t)$ plane in Fig.\ref{fig:t-W_c} for charmonium and in Fig.\ref{fig:t-W_b} for bottomonium.

In the near threshold region $s\gtrsim (M_N+M_X)^2$, the factorization on the proton side only works when the outgoing meson is heavy enough, such that the proton target moves fast enough to be factorized using the parton picture. In the heavy meson limit, the incoming and outgoing nucleon velocity is of order 1 up to some correction proportional to the mass ratio $M_N^2/M_X^2$. Therefore, the factorization method  using the parton description  still holds near the photoproduction threshold. On the other hand, near the threshold region, there is not much energy left to move the heavy meson. The outgoing meson velocity becomes non-relativistic. Therefore, the meson part can be analysed using  non-relativistic QCD (NRQCD). Close to threshold, the skewness $\xi$ is close to one. Similar arguments for the photoproduction near threshold can also be found in \cite{Guo:2021ibg,Sun:2021pyw} for $J/\psi$ and in \cite{Ma:2003py} for $\eta_c$.

\subsection{$J/\psi$ meson photoproduction}
For a fast moving hadron, the constituent becomes collinear. In the heavy limit, the dominant process of the scattering in Fig.\ref{fig:scattering} can be factorized into a hard scattering with outgoing partons carrying momentum $k^+_i$ and the parton correlation in the nucleon. Thus, the factorization amplitude of the photoproduction near threshold can be written as
\begin{widetext}
\bea
    i\mathcal{M}(\gamma N\rightarrow J/\psi\ N')
    =&&\int_{-P^+}^{\infty} dk^+_1\int_{-P^+}^{\infty}dk^+_2W^{ab}_{\mu\nu}(k_1^+,k_2^+)\frac{i}{k_1^++i0^+}\frac{i}{k_2^++i0^+}\nonumber\\
    &&\times\int \frac{d\lambda_1^-}{2\pi}\frac{d\lambda_2^-}{2\pi}e^{-ik^+_1\lambda_1^--ik^+_2\lambda_2^-}\langle N'|F^{a\mu+}\left(\lambda_2^-\right) F^{b\nu+}\left(\lambda_1^-\right)|N\rangle
\eea
\end{widetext}
We use the light-cone gauge,  where the unphysical longitudinally polarized collinear gluons vanish. The lower bound of the parton momentum-integral assures that the spectators in the nucleon remains physical, when the parton comes out from the nucleon.

\begin{figure}
\subfloat[\label{fig_J1}]{%
  \includegraphics[scale=0.7]{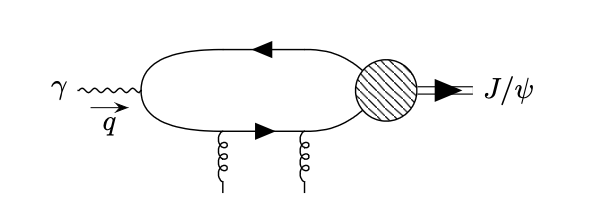}%
}\hfill
\subfloat[\label{fig_J2}]{%
  \includegraphics[scale=0.7]{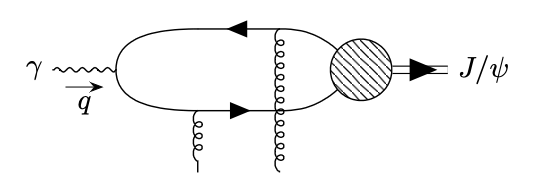}%
}
\caption{Feynman diagram for the hard kernel in $J/\psi$ photoproduction.}
\label{fig_Jpsi}
\end{figure}
The hard kernel does not depend on $\Delta^+$ due to translational symmetry in the theory. The relative momentum $k^+$ dependence can be computed using the Feynman diagrams in Fig. \ref{fig_Jpsi}. The transversely polarized outgoing quarkonium dominates the amplitude in the heavy meson limit \cite{Sun:2021pyw}. Thus, the hard kernel reads
\begin{equation}
    W^{ab}_{\mu\nu}=\frac{g^2}{2}\frac{\delta^{ab}g_{\perp\mu\nu}}{\sqrt{N_c}}\frac{\psi^*_{J/\psi}(0)}{m_c}(8\epsilon_\gamma\cdot\epsilon_V^*)
\end{equation}
where $\psi_{J/\psi}$ is the non-relativistic wave function defined in Appendix \ref{Appx:heavy_meson_WF}. 
In the heavy meson photoproduction process, the mesonic part in the hard kernels are treated by non-relativistic QCD. The relative momentum between the quark-antiquark pair is of order $\mathcal{O}(\alpha_sM_X)$, and the heavy meson mass is assumed to be $M_X=2m_Q$. 

To simplify the amplitude, we use the  symmetric parameterization
\begin{align*}
\lambda_c^-=&\frac{\lambda_1^-+\lambda_2^-}{2}  & \Delta^+=&k^+_1+k_2^+=-2\xi\bar{P}^+ \\
\lambda^-=&\lambda_1^--\lambda_2^- & k^+=&\frac{k^+_1-k_2^+}{2}=x\bar{P}^+
\end{align*}
Translational symmetry implies no dependence on $\lambda_c^-$ in the parton correlation. With this in mind, the photoproduction amplitude reads
\begin{widetext}
\begin{equation}
\begin{aligned}
    &i\mathcal{M}(\gamma N\rightarrow J/\psi\ N')= \frac{g^2}{\sqrt{N_c}}\frac{\psi^*_{J/\psi}(0)}{m_c}(4\epsilon_\gamma\cdot\epsilon_V^*)\mathcal{W}_{2g}(t,\xi)
\end{aligned}
\end{equation}
Here $\mathcal{W}_{2g}(t,\xi)$  depends implicitly on the polarization of the initial and final proton, through
\begin{equation}
    \mathcal{W}_{2g}(t,\xi)=\int_{-1}^{1} dx\frac{1}{x-\xi+i0^+}\frac{1}{x+\xi-i0^+}f_{2g}(x,t,\xi)
\end{equation}
where the two-gluon distribution is defined as
\begin{equation}
    f_{2g}(x,\xi,t)=\int \frac{d\lambda^-}{2\pi}e^{-ix\bar{P}^+\lambda^-}\frac{1}{\bar{P}^+}\langle N'|F^{a+i}\left(-\lambda^-/2\right) F^{a+}{}_i\left(\lambda^-/2\right)|N\rangle
\end{equation}
\end{widetext}

\subsection{$\eta_c$ meson photoproduction}
The threshold photoproduction analysis of heavy pseudoscalars
follow a similar reasoning. The dominant process in the heavy meson limit, is factorized into a hard scattering with three outgoing gluons carrying momenta $k^+_1, k_2^+, k^+_3$, times the three-gluon light-front correlation in the nucleon,
\begin{widetext}
\begin{align}
i\mathcal{M}(\gamma N\rightarrow \eta_c\ N')=&\int^{\infty}_{-P^+} dk^+_1\int^{\infty}_{-P^+}dk^+_2\int^{\infty}_{-P^+}dk^+_3W^{abc}_{\mu\nu\sigma}(k_1^+,k_2^+,k_3^+)\frac{i}{k_1^++i0^+}\frac{i}{k_2^++i0^+}\frac{i}{k_3^++i0^+}\nonumber\\
    &\times\int\frac{d\lambda_1^-}{2\pi}\frac{d\lambda_2^-}{2\pi}\frac{d\lambda_3^-}{2\pi}e^{-ik^+_1\lambda_1^--ik^+_2\lambda_2^--ik^+_3\lambda_3^-}\langle N'|F^{a\mu+}\left(\lambda_3^-\right) F^{b\nu+}\left(\lambda_2^-\right)F^{c\sigma+}\left(\lambda_1^-\right)|N\rangle
\end{align}
\end{widetext}
The  light-cone gauge is used again to remove the unphysical longitudinally polarized collinear gluons. The lower bound of the parton momentum-integral assures the spectators in the nucleon remains physical when the parton comes out from the nucleon.

\begin{figure}
\subfloat[\label{fig_ec1}]{%
  \includegraphics[scale=0.6]{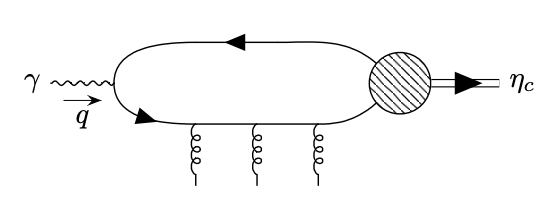}%
}\hfill
\subfloat[\label{fig_ec2}]{%
  \includegraphics[scale=0.6]{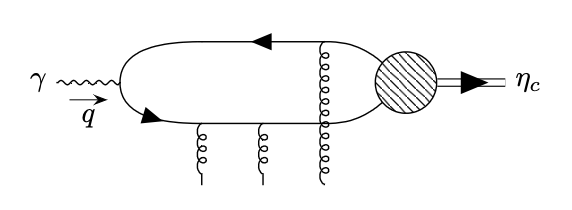}%
}
\caption{Feynman diagram for the hard kernel for $\eta_c$ photoproduction.}
\label{fig_etac}
\end{figure}

Again, the hard kernel $W^{abc}_{\mu\nu\sigma}$ does not depend on $\Delta^+$ due to translational symmetry. The relative momenta $k_\lambda^+$ and $k^+_\rho$ dependence can be computed using the Feynman diagrams in Fig. \ref{fig_etac}. In the heavy meson limit, the hard kernel reads
\begin{widetext}
\begin{equation}
\begin{aligned}
W^{abc}_{\mu\nu\sigma}=&\frac{ig^3}{2}d^{abc}\epsilon_{\mu-+\alpha}\epsilon^\alpha_{\gamma}(q)\frac{g_{\perp\nu\sigma}}{\sqrt{N_c}}\frac{\psi^*_{\eta_c}(0)}{m_c}\\
&\times\left[4\sqrt{2}\left(\frac{2i}{k_1^++k_2^++i0^+}+\frac{i}{k_1^++k_3^++i0^+}+\frac{i}{k_2^++k_3^++i0^+}\right)\right]    
\end{aligned}
\end{equation}
where $\psi_{\eta_c}$ is the non-relativistic wave function of the $\eta_c$ meson defined in Appendix \ref{Appx:heavy_meson_WF}. 
To further simplify the amplitude, we use the symmetric parameterization
\begin{align*}
\lambda_c^-=&\frac{\lambda_1^-+\lambda_2^-+\lambda_3^-}{3}  & 
\lambda^-=&\lambda_1^--\lambda_2^- & 
\rho^-=&\frac{1}{2}\left(\lambda_1^-+\lambda_2^--2\lambda^-_3\right) \\
\Delta^+=&k^+_1+k_2^++k_3^+=-2\xi\bar{P}^+
 & k_\lambda^+=&\frac{k^+_1-k_2^+}{2}=x_\lambda\bar{P}^+
 & k_\rho^+=&\frac{k^+_1+k_2^+-2k_3^+}{3}=x_\rho\bar{P}^+
\end{align*}
The dependence on $\lambda_c^-$ drops out from the parton correlation because of translational symmetry, hence
\begin{equation}
\begin{aligned}
    &i\mathcal{M}(\gamma N\rightarrow \eta_c\ N')=-\frac{ig^3}{\sqrt{N_c}}\frac{\psi^*_{\eta_c}(0)}{m_c}2\sqrt{2}\epsilon_{\perp ij}\epsilon_{\gamma}^i(q)\mathcal{W}^j_{3g}(t,\xi)
\end{aligned}
\end{equation}
The Levi-Civita tensor in the transverse space is defined as $\epsilon_{\perp ij}=\epsilon_{0ij3}$ with the convention $\epsilon_{0123}=1$. The transverse vector function $\mathcal{W}^i_{3g}(t,\xi)$ depends implicitly on the polarization of the initial and final proton, and is defined as
\begin{equation}
\begin{aligned}
    \mathcal{W}^i_{3g}(t,\xi)=&\int_{-2-\xi/3}^{1+\xi/3} dx_\rho\int_{-1-\xi/3-x_\rho/2}^{1+\xi/3+x_\rho/2} dx_\lambda\\
    &\times\left(\frac{2}{\frac{4}{3}\xi-x_\lambda-i0^+}+\frac{1}{\frac{4}{3}\xi-x_\rho+\frac{1}{2}x_\lambda-i0^+}+\frac{1}{\frac{4}{3}\xi+x_\rho+\frac{1}{2}x_\lambda-i0^+}\right)\\
    &\times\bigg(\frac{1}{\frac{2}{3}\xi-x_\lambda-\frac{1}{2}x_\rho-i0^+}\frac{1}{\frac{2}{3}\xi+x_\lambda-\frac{1}{2}x_\rho-i0^+}\frac{1}{\frac{2}{3}\xi+x_\rho-i0^+}f^i_{3g}(x_\rho,x_\lambda,\xi,t)\bigg)
\end{aligned}
\end{equation}
The three-gluon distribution is defined as
\begin{equation}
\begin{aligned}
    f^i_{3g}(x_\rho,x_\lambda,\xi,t)=&\int \frac{d\lambda^-}{2\pi}\int \frac{d\rho^-}{2\pi}e^{-ix_\lambda\bar{P}^+\lambda^--ix_\rho\bar{P}^+\rho^-}\frac{d^{abc}}{(\bar{P}^+)^2}\\
    &\times\langle N'|F^{a+i}\left(-\frac{2}{3}\rho^-\right) F^{b+j}\left(\frac{1}{2}\lambda^-+\frac{1}{3}\rho^-\right)F^{c+}{}_j\left(\frac{1}{2}\lambda^-+\frac{1}{3}\rho^-\right)|N\rangle
    \end{aligned}
\end{equation}
\end{widetext}

\subsection{Large skewness expansion}
In the threshold region, the exchanged gluons in the factorized proton matrix elements can be organized using a twist expansion (See Appendix \ref{Appx:OPE}). Each of the local operators in the operator product expansion (OPE),  sources a colorless gluonic exchange in the form of a glueball. The corresponding glueball exchanges, can be organized using charge conjugation and parity. However, parity changes the direction of spatial momentum and flip the helicity. Parity symmetry on the light front has to be combined with time reversal, so that the light front direction is  unchanged. We adopt the $\Lambda$-parity in \cite{Ji:2003yj} by making an additional $180^{\circ}$ rotation around the $y$-axis to restore the longitudinal momentum, i.e. $ Y= e^{-i\pi J_y}P$.
As a result, the plus component is transformed in the same way as the time component under the usual parity symmetry.

With this in min, charge conjugation and parity symmetry, show  that the OPE for $f_{2g}(x,\xi,t)$ is dominated by a $2^{++}$ glueball operator,  and the OPE of $f^i_{3g}(x_\rho,x_\lambda,\xi,t)$ by a  $1^{+-}$ glueball operator.
Among these operators with the same twist, operators with higher dimension corresponds to higher glueball excitations (Regge spectrum). Since there is not much energy to excite these gluebalsl near threshold, only the lowest dimensional operator, which represents the ground state of the glueball dominates the non-perturbative nucleon matrix element. Therefore, the non-local gluonic operators can be approximately treated as a local operator
\begin{widetext}
\begin{equation}
\begin{aligned}
F^{a+i}\left(-\lambda^-/2\right) F^{a+}{}_i\left(\lambda^-/2\right)\simeq F^{a+i}F^{a+}{}_i(0)
\end{aligned}
\end{equation}
\begin{equation}
\begin{aligned}
    d^{abc}F^{a+\mu}\left(-\frac{2}{3}\rho^-\right) F^{b+i}\left(-\frac{1}{2}\lambda^-+\frac{1}{3}\rho^-\right)F^{c+}{}_i\left(\frac{1}{2}\lambda^-+\frac{1}{3}\rho^-\right)
    \simeq d^{abc}F^{a+\mu} F^{b+i}F^{c+}{}_i(0)
\end{aligned}
\end{equation}
In the local approximation and the large skewness $\xi$ limit, the functions $\mathcal{W}_{2g}$ and $\mathcal{W}^\mu_{3g}$ in  the leading moment are
\begin{equation}
   \mathcal{W}_{2g}(t,\xi\rightarrow1)=\frac{1}{\xi^2(\bar{P}^+)^2}\langle P'|F^{a+i} F^{a+}{}_i|P\rangle
\end{equation}
\begin{equation}
   \mathcal{W}^\mu_{3g}(t,\xi\rightarrow1)=\left(\frac{3}{2}\right)^3\frac{3}{\xi^4(\bar{P}^+)^4} d^{abc}\langle P'|F^{a+\mu}F^{b+i}F^{c+}{}_i|P\rangle
\end{equation}


\subsection{Two-gluon distribution}
The threshold photoproduction of $J/\Psi$ is dominated by the $C$-even $2^{++}$ glueball exchange,  parameterized by the gluon generalized parton distribution (GPD)

\begin{equation}
\begin{aligned}
    f_{2g}(x,\xi,t)
    =\frac{1}{\bar{P}^+}\bar{u}_{s'}(P')\left[xH_g(x,\xi,t)\gamma^++xE_g(x,\xi,t)\frac{i\sigma^{+\alpha}\Delta_\alpha}{2M_N}\right]u_{s}(P)
\end{aligned}
\end{equation}
\end{widetext}
with  $H_g(x,\xi,t), E_g(x,\xi,t)$ the gluon GPDs \cite{Ji:1998pc}. 
The zeroth moment of the gluon distribution dominates the process near  threshold. As the zeroth moment of $f_{2g}(x,\xi,t)$ is related to the nucleon matrix element of the energy momentum tensor, the matrix element can be parameterized by 
\begin{widetext}
\begin{equation}
\begin{aligned}
\label{eq:f2g}
    \int_{-1}^1dx f_{2g}(x,\xi,t)=\frac{1}{(\bar{P}^+)^2}\langle P'|F^{a+ i} F^{a+}{}_i|P\rangle
    =\frac{1}{\bar{P}^+}\bar{u}_{s'}(P')\left[H_{2g}(t,\xi)\gamma^++E_{2g}(t,\xi)\frac{i\sigma^{+\alpha}\Delta_\alpha}{2M_N}\right]u_s(P)
\end{aligned}
\end{equation}
\end{widetext}
These twist-$2$ functions $H_{2g}(t,\xi)$ and $E_{2g}(t,\xi)$ are related to the second Mellin moment of the gluon GPD. Hermicity and time reversal symmetry imply that the twist-$2$ functions are even functions of the skewness $\xi$, with 
\bea
H_{2g}(t,-\xi)&=&H_{2g}(t,\xi)\nonumber\\
E_{2g}(t,-\xi)&=&E_{2g}(t,\xi)
\eea

\subsection{Three-gluon distribution}
The threshold photoproduction of heavy pseudoscalars is dominated by $C$-odd $1^{+-}$ glueball exchanges.
Following~\cite{Ma:2003py,Guo:2021aik}, the three-gluon distribution can be parameterized by the four twist-3 gluon GPDs $G_{g1}(x_\rho,x_\lambda,t,\xi)$, $G_{g2}(x_\rho,x_\lambda,t,\xi)$, $G_{g3}(x_\rho,x_\lambda,t,\xi)$ and $G_{g4}(x_\rho,x_\lambda,t,\xi)$. The parameterization is similar to the generalized helicity flip quark distribution in \cite{Diehl:2001pm}
\begin{widetext}
\begin{equation}
\begin{aligned}
    f^i_{3g}(x_\rho,x_\lambda,\xi,t)=\frac{M_N}{(\bar{P}^+)^2}\bar{u}_{s'}(P')\left[G_{g1}\sigma^{+i}+G_{g2}\frac{i\bar{P}^+\Delta^i}{M_N^2}+G_{g3}\frac{i\Delta^i\gamma^+}{M_N}+G_{g4}\frac{\bar{P}^+\sigma^{i\alpha}\Delta_\alpha}{2M_N^2}\right]u_s(P)
\end{aligned}
\end{equation}
\end{widetext}
where we abbreviate the notation of twist-3 gluon GPDs for simplicity. The zeroth moment of the gluon distribution dominates the process near  threshold. Since the zeroth moment of $f^i_{3g}(x,\xi,t)$ is related to the nucleon matrix element of the $C$-odd three gluon operator, the matrix element can be parameterized by 
\begin{widetext}
\begin{equation}
\label{eq:odd_gluball1}
\begin{aligned}
    &\int_{-2-\xi/3}^{1+\xi/3} dx_\rho\int_{-1-\xi/3-x_\rho/2}^{1+\xi/3+x_\rho/2} dx_\lambda f^i_{3g}(x_\rho,x_\lambda,\xi,t)=\frac{1}{(\bar{P}^+)^4} d^{abc}\langle P'|F^{a+i}F^{b+j}F^{c+}{}_j|P\rangle\\
    =&\frac{M_N}{(\bar{P}^+)^2}\bar{u}_{s'}(P')\left[f_1(t,\xi)\sigma^{+i}+f_2(t,\xi)\frac{i\bar{P}^+\Delta^i}{M_N^2}+f_3(t,\xi)\frac{i\Delta^i\gamma^+}{M_N}+f_4(t,\xi)\frac{\bar{P}^+\sigma^{i\alpha}\Delta_\alpha}{2M_N^2}\right]u_s(P)
\end{aligned}
\end{equation}
\end{widetext}
The twist-$3$ functions $f_{i}(t,\xi)$ are related to the zeroth moment of the $F$-type twist-3 gluon GPDs $G_{gi}(x_\rho x_\lambda,\xi,t)$
and all proportional to $M_N/\bar{P}^+$ since they correspond to twist-3 operators. The hermicity of the matrix element implies
\bea
f_{1,2,3}(t,-\xi)&=&f_{1,2,3}(t,\xi)\nonumber\\
f_4(t,-\xi)&=&-f_4(t,\xi)
\eea
In our evaluations to follow, these twist-$3$ functions vanish in forward limit $\xi\rightarrow0$. They are related to the magnetic moment form factor $f_g(t)$. Higher order terms corresponds to other form factors that are induced by different resonance in the glueball exchange channels. 

In the limit $\Delta^i_\perp\rightarrow0$ near threshold, $f_2$ and $f_3$ vanish while $f_1$ and $f_4$ do not. Since the initial photon has the helicity $\pm1$, and $\eta_c$ has the helicity $0$, the proton has to flip its total angular momentum in the $z$-direction by $\pm1$ in order to satisfy  angular momentum conservation.
This change can be fulfilled by the change of the orbital angular momentum, whose effects are captured by $\Delta^i_\perp$ in \eqref{eq:odd_gluball1}, or made by changing the helicity of the proton, which is denoted by the $z$-component of the proton spin $s_z$. It is clear that in the limit $\Delta^i_\perp\rightarrow0$, the orbital contributions parameterized by $f_2$ and $f_3$ become zero. $f_1$ corresponds to the chiral flipping process, and $f_4$ can be mapped to the chiral conserving process by $$\bar{u}_s(P')\frac{\sigma^{i\alpha}\Delta_\alpha}{2M_N}u_s(P)\bigg|_{\Delta^i_\perp=0}=-\bar{u}_s(P')i\gamma^i_\perp u_s(P)$$ 
Therefore, the contribution comes from the helicity flipping, which corresponds to the chirality flipping term $f_1$ and chirality conserving $f_4$. In the massless limit of the proton, the chirality is exactly equal to helicity. The nonzero contribution of $f_4$ in the forward limit is due to the finite mass correction between the helicity and chirality.

\section{Gluonic form factors}
\label{SEC3}
At low momentum transfer, the C-even and C-odd gluonic matrix elements in the proton state, can be evaluated in the QCD instanton vacuum.  In Appendix~\ref{SEC3X}, we briefly outline the derivation, with most details to be presented in~\cite{Liu:2024}.

\subsection{C-even gluonic operator}
The dominant C-even gluonic operator in the threshold photoproduction of $J/\Psi$ is 
\begin{equation}
\label{eq:O2g}
    \mathcal{O}^{++}_{2g}[A]=F^{a+i}(x) F^{a+}{}_i(x)
\end{equation}
It is related to the gluonic energy momentum tensor $T_g^{++}$, with the matrix element in a nucleon state
\begin{widetext}
\begin{equation}
\begin{aligned}
\label{eq:matrix_2g}
\langle P'|F^{a+ i} F^{a+}{}_i|P\rangle
=\bar{u}_{s'}(P')\bigg[A_g(t)\gamma^{+}\bar{P}^{+}+B_g(t)\frac{i\bar{P}^{+}\sigma^{+\alpha}\Delta_\alpha}{2M_N}+C_g(t)\frac{(\Delta^+)^2}{M_N}\bigg]u_s(P)   
\end{aligned}
\end{equation}
\end{widetext}
Note that in light front signature (\ref{eq:O2g})reads 
\begin{equation}
\begin{aligned}
   &F^{a+i}F^{a+}{}_{i}=\\
    &-\frac{1}{2}\left(\vec{E}_\perp^a\cdot \vec{E}_\perp^a+2\hat{z}\cdot(\vec{E}_\perp^a\times \vec{B}_\perp^a)+\vec{B}_\perp^a\cdot \vec{B}_\perp^a\right)
\end{aligned}
\end{equation}
Since the pseudoparticles are self-dual tunneling configurations with $\vec{E}^a=\mp i\vec{B}^a$,
the leading contribution in the pseudoparticle density vanishes. The non-vanishing contributions start from the pair correlations illustrated in Fig.~\ref{fig:inst_op}, at next to leading order in the density~\cite{Zahed:2021fxk}. 

With this in mind, and inserting (\ref{eq:O2g}) in the ensemble averaging (\ref{eq:o_had_exp}) yields~\cite{Liu:2024}
\begin{widetext}
\begin{equation}
\begin{aligned}
\label{eq:A_g}
    \langle P'S|g^2\mathcal{O}^{++}_{2g}[A]|PS\rangle=&\frac{4\kappa^2}{2N_c(N_c^2-1)}J_{IA}(\rho m^*)\Bigg\{\frac{16\pi^2}{3}\beta^{(IA)}_{T_g,1}(\rho \sqrt{-t})\langle P'S|\bar{\psi}\gamma^{+} i\overleftrightarrow{\partial}^{+}\psi|PS\rangle\\
    &-\frac{4\pi^2\rho^2}{9}\beta^{(IA)}_{T_g,2}(\rho \sqrt{-t})\Delta^+\Delta_\lambda\langle P'S|\bar{\psi}\left(\gamma^{(+}i\overleftrightarrow{\partial}^{\lambda)}-\frac{1}{4}g^{+\lambda}i\overleftrightarrow{\slashed{\partial}}\right)\psi|PS\rangle\\
    &-4\pi^2\rho^4\beta^{(IA)}_{T_g,3}(\rho \sqrt{-t})\left(\Delta^+\right)^2\Delta^\rho \Delta^\lambda\langle P'S|\bar{\psi}\left(\gamma_{(\rho}i\overleftrightarrow{\partial}_{\lambda)}-\frac{1}{4}g_{\rho\lambda}i\overleftrightarrow{\slashed{\partial}}\right)\psi|PS\rangle\Bigg\}\\
    &+\frac{4\kappa^2}{2N_c(N_c^2-1)}J_{II}(\rho m^*)\frac{8\pi^2\rho^2}{9}\beta^{(II)}_{T_g}(\rho \sqrt{-t})\left(\Delta^+\right)^2\frac{1}{\rho}\langle P'S|\bar{\psi}\psi|PS\rangle
\end{aligned}
\end{equation}
The induced form factors by the pseudoparticles finite size are defined as
\begin{equation}
    \beta^{(IA)}_{T_g,1}(q)=\frac{1}{q}\int_0^\infty dx\left[\frac{24}{(1+x^2)^4}J_1(qx)+\frac{24x^2}{(1+x^2)^4}J_3(q x)-\frac{192}{(1+x^2)^3}\frac{J_3(qx)}{q^2x^2}\right]
\end{equation}

\begin{equation}
    \beta^{(IA)}_{T_g,2}(q)=\frac{1}{q}\int_0^\infty dx9x^2\left[\frac{128x^2}{(1+x^2)^4}\frac{J_3(qx)}{q^2x^2}-\frac{512}{(1+x^2)^3}\frac{J_4(qx)}{q^3x^3}\right]
\end{equation}

\begin{equation}
    \beta^{(IA)}_{T_g,3}(q)=\frac{1}{q}\int_0^\infty dx\frac{256x^4}{(1+x^2)^3}\frac{J_5(qx)}{q^4x^4}
\end{equation}

\begin{equation}
    \beta^{(II)}_{T_g}(q)=\frac{1}{q}\int_0^\infty dx\frac{576x^2}{(1+x^2)^4}\frac{J_3(qx)}{q^2x^2}
\end{equation}
Recall that the quark energy-momentum of the nucleon are parameterized  by the gravitational form factors $A_q(t)$, $B_q(t)$, $C_q(t)$ defined as \cite{Ji:2000id,Polyakov:2019lbq}
\begin{equation}
\begin{aligned}
\label{eq:quark_2g}
&\langle P'|\bar{\psi}\left(\gamma^{(\mu} i\overleftrightarrow{\partial}^{\nu)}-\frac{1}{4}g^{\mu\nu}i\overleftrightarrow{\slashed{\partial}}\right)\psi|P\rangle
\\
=&\bar{u}_{s'}(P')\bigg[A_q(t)\left(\gamma^{(\mu} \bar{P}^{\nu)}-\frac{1}{4}g^{\mu\nu}M_N\right)\\
&+B_q(t)\left(\frac{i\bar{P}^{(\mu}\sigma^{\nu)\alpha}\Delta_\alpha}{2M_N}-g^{\mu\nu}\frac{\Delta^2}{16M_N}\right)+C_q(t)\frac{\Delta^{\mu}\Delta^{\nu}-\frac{1}{4}g^{\mu\nu}\Delta^2}{M_N}\bigg]u_s(P)   
\end{aligned}
\end{equation}
Also, the nucleon sigma term is given by
\begin{equation}
    \langle P'|\bar{\psi}\psi|P\rangle=\frac{\sigma(t)}{m}\bar{u}_s(P')u_s(P)
\end{equation}
A comparison of \eqref{eq:quark_2g} with \eqref{eq:matrix_2g}, 
ties the gluon and quark gravitational form factors at next to  
leading order in the pseudoparticle density in the QCD instanton vacuum,
\begin{equation}
\label{AGTX}
\begin{aligned}
    A_g(t)=&\frac{1}{2N_c(N_c^2-1)}4\kappa^2J_{IA}(\rho m^*)\frac{16\pi^2}{3}\beta^{(IA)}_{T_g,1}(\rho \sqrt{-t})A_q(t)
\end{aligned}
\end{equation}
\begin{equation}
\begin{aligned}
\label{eq:B_g}
    B_g(t)=&\frac{1}{2N_c(N_c^2-1)}4\kappa^2J_{IA}(\rho m^*)\frac{16\pi^2}{3}\beta^{(IA)}_{T_g,1}(\rho \sqrt{-t})B_q(t)
\end{aligned}
\end{equation}
\begin{equation}
\label{CTXX}
\begin{aligned}
    &C_g(t)=\frac{1}{2N_c(N_c^2-1)}4\kappa^2J_{IA}(\rho m^*)\\
    &\times\Bigg\{\frac{16\pi^2}{3}\beta^{(IA)}_{T_g,1}(\rho \sqrt{-t})C_q(t)\\
    &\qquad+\frac{2\pi^2\rho^2}{9}M_N^2\left(\beta^{(IA)}_{T_g,2}(\rho \sqrt{-t})+\frac{9}{2}\rho^2t\beta^{(IA)}_{T_g,3}(\rho \sqrt{-t})\right)\left(A_q+B_q\frac{t}{4M^2_N}-C_q\frac{3t}{M^2_N}\right)\Bigg\}\\
    &\qquad+\frac{1}{2N_c(N_c^2-1)}4\kappa^2J_{II}(\rho m^*)\frac{4\pi^2}{9}\rho M_N\beta^{(II)}_{T_g}(\rho \sqrt{-t})\frac{\sigma(t)}{m} 
\end{aligned}
\end{equation}
\end{widetext}

In the forward limit, the gluonic charge $A_g(0)$ can be evaluated by noting that $A_g(0)+A_q(0)=1$, which allows for solving for $A_g(0)$ using (\ref{AGTX}). Using the QCD instanton vacuum parameters and DGLAP evolution, 
the result at the resolution of $\mu=2{\rm GeV}$ is~\cite{Liu:2024}
\bea
\label{A00}
A_g(0)=0.391
\eea
This value is close to the gluon momentum prediction  $\langle x\rangle_g=0.414$ following from the CTEQ global analysis in~\cite{Hou:2019efy,Guo:2023pqw}.  It is also consistent with the lattice result  $A_g(0)=0.501$~\cite{Hackett:2023rif}. 
The details for the evaluation of the charge $C_{g}(0)$ following from (\ref{CTXX}),
and using the QCD instanton vacuum parameters and DGLAP evolution are given in~\cite{Liu:2024}, with the result at the resolution of $\mu=2{\rm GeV}$
\bea
\label{C00}
C_g(0)=-0.4563
\eea

\begin{figure*}
\label{FFACF}
\subfloat[\label{fig:Ag(t)}]{%
\includegraphics[scale=0.52]{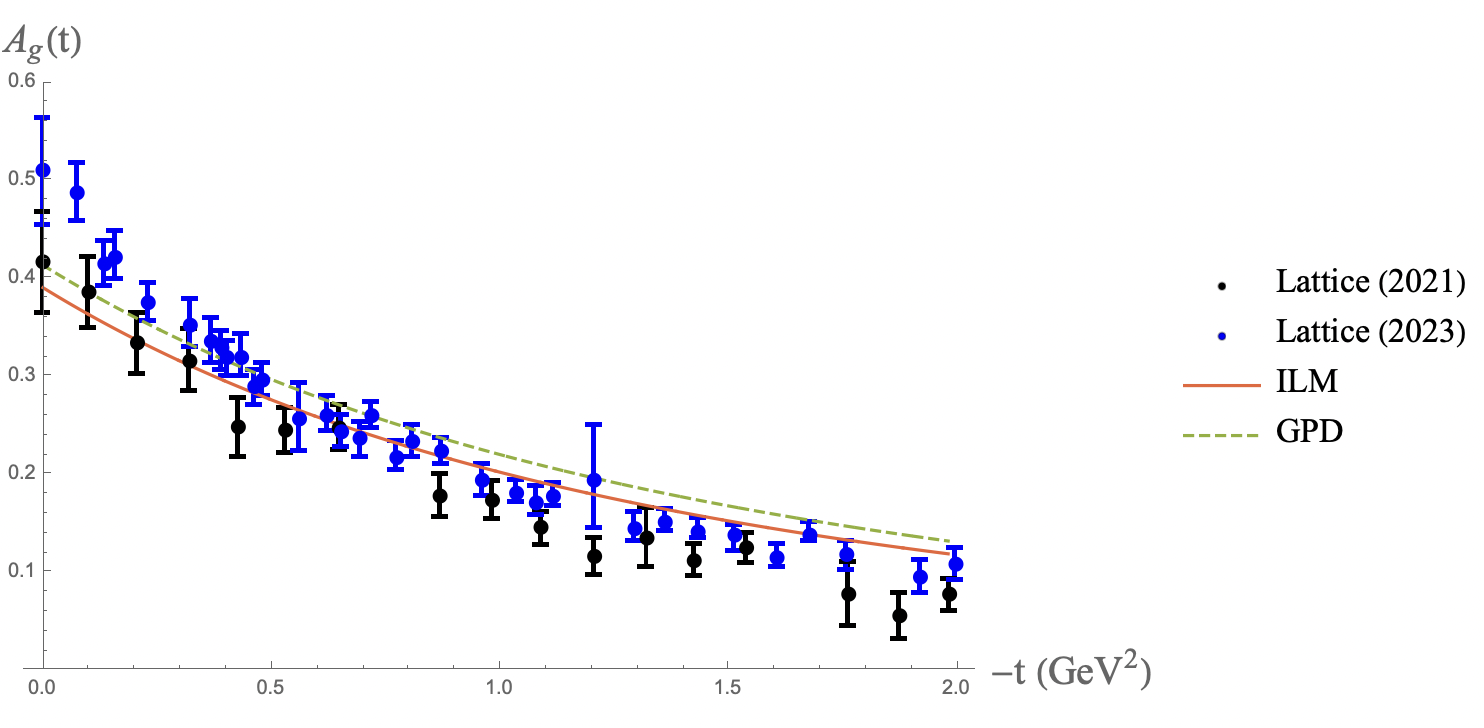}%
}\hfill
\subfloat[\label{fig:Cg(t)}]{%
\includegraphics[scale=0.52]{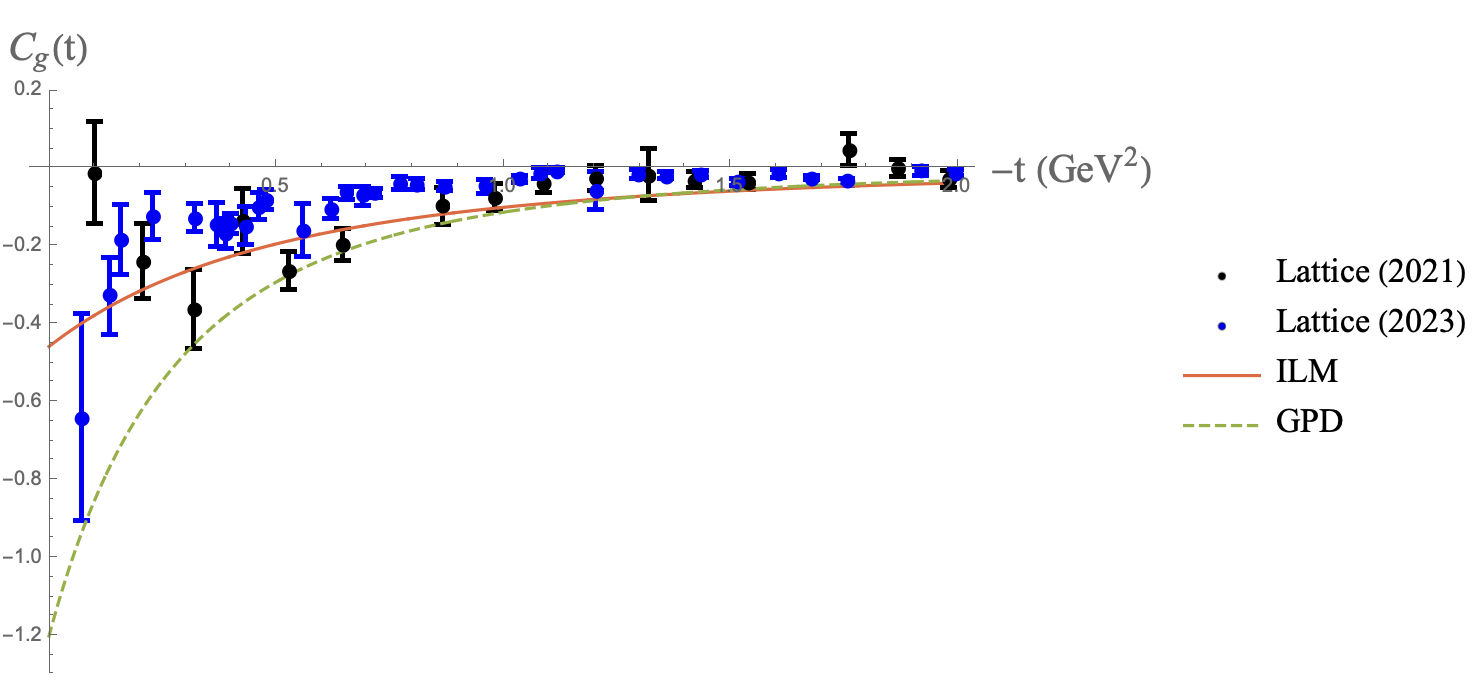}%
}\hfill
\subfloat[\label{fig:fg(t)}]{%
\includegraphics[scale=0.55]{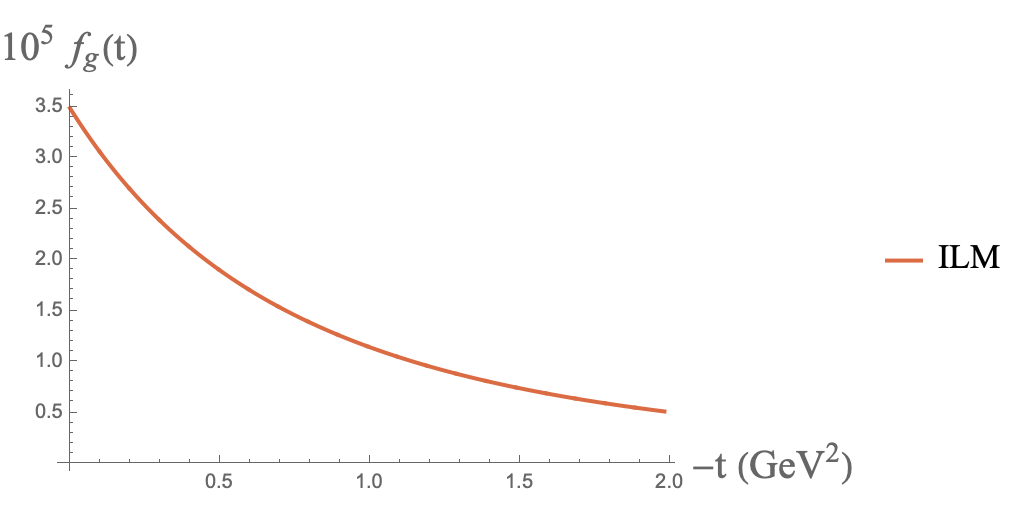}%
}
\caption{ a,b: The instanton estimation of the gluonic gravitational form factors $A_g, C_g$ in (\ref{ACF}) (orange-solid curves) compared to lattice results
 with a large pion mass $m_\pi=450$ MeV (black data points) \cite{Pefkou:2021fni},  and  those with a smaller pion mass $m_\pi=170$ MeV (blue data points) \cite{Hackett:2023rif};
c: The instanton estimation of the C-odd gluonic form factor in (\ref{ACF}).
}
\end{figure*}

\subsection{C-odd gluon operator}
Similarly to the C-even gluonic operator for threshold photoproduction of $J/\Psi$, the C-odd gluonic operator 
\begin{equation}
\label{3g+++}
    \mathcal{O}^{+++i}_{3g}[A]=d^{abc}F^{a+i}(x)F^{b+j}(x)F^{c+}{}_j(x)
\end{equation}
is dominant in the photoproduction of heavy pseudoscalar mesons
such as $\eta_{c,b}$~\cite{Ma:2003py}. 

To proceed with the ensemble average of (\ref{3g+++}) in the QCD instanton vacuum, we note that the symmetric $SU(3)$ structure constant $d^{abc}$ has no support in the  $SU(2)$ subgroup. 
Hence, individual $SU(2)$ pseudoparticles cannot contribute in leading order, even if embedded in $SU(N)$ since
$\mathrm{tr}(\tau^a\{\tau^b,\tau^c\})=0$. The contributions stem
from pairs of pseudoparticles at next-to-leading order, with the result~\cite{Liu:2024}
\begin{widetext}
\begin{equation}
\begin{aligned}
\label{eq:had_3g}
    &\langle P'S|g^3\mathcal{O}^{+++i}_{3g}[A]|PS\rangle=\frac{N_c-2}{8N_c^2(N_c^2-1)(N_c+2)}4\kappa^2J_{II}(\rho m^*)\\
    &\times(\Delta^+)^2\frac{2\pi^2\rho^2}{225}\beta^+_{3g}(\rho\sqrt{-t})\frac{1}{\rho}\langle P'S|\bar{\psi}\left(\sigma^{i\alpha}\Delta^+\Delta_\alpha-\sigma^{+\alpha}\Delta^i\Delta_\alpha-\frac{1}{2}\Delta^2\sigma^{+i}\right)\psi|PS\rangle\\
\end{aligned}
\end{equation}
\end{widetext}
where
\begin{equation}
    \beta^+_{3g}(q)=\frac{450}{q}\int_0^\infty dx\frac{512x^4}{(1+x^2)^6}\frac{J_5(qx)}{q^4x^4}
\end{equation}
The quark tensor matrix element is related to the nucleon tensor charge $\delta q(\mu)$,
\begin{equation}
\label{eq:tensor}
    \langle P'S|\bar{\psi}\sigma^{\mu\nu}\psi|PS\rangle=\delta q(t) \bar{u}_{s}(P')\sigma^{\mu\nu}u_{s}(P)
\end{equation}  
From (\ref{eq:had_3g}) it follows that 
\begin{widetext}
\begin{equation}
\begin{aligned}
\label{TWIST3X}
    &\langle P'|d^{abc}F^{a+ i}F^{b+ j} F^{c+}{}_j|P\rangle\\
    &=f_g(t)\bar{u}_s(P')\frac{1}{2M_N}\left(\sigma^{i\alpha}\Delta_\alpha(\Delta^+)^3-\sigma^{+\alpha}\Delta_\alpha\Delta^i(\Delta^+)^2-\frac{1}{2}\Delta^2\sigma^{+i}(\Delta^+)^2\right)u_s(P)
    \end{aligned}
\end{equation}
with the induced gluon magnetic form factor
\begin{equation}
\label{TWIST3XX}
    f_g(t)=\frac{N_c-2}{4N_c^2(N_c^2-1)(N_c+2)}4\kappa^2J_{II}(\rho m^*)\frac{2\pi^2}{225}\rho M_N\beta^{+}_{3g}(\rho\sqrt{-t})\delta q(t)
\end{equation}
\end{widetext}

The value of the nucleon gluonic charge $f_g(0)$ is tied to the nucleon tensor charge 
$\delta q(0)$. The latter  can be extracted from the experimental data obtained in the back-to-back di-hadron productions in $e^+e^-$ annihilation, with the result $\delta q(0)=0.185$ at $\mu=1.55$ GeV~\cite{YE201791,Kang:2015msa}. Although the value changes with the renormalization scale $\mu$ due to the anomalous dimensions of the tensor charge~\cite{artru1990transversely,Gamberg:2001xc}, the evolution is slow with
the result at $\mu=2\,{\rm GeV}$~\cite{Liu:2024}
\bea
\label{F00}
f_g(0)=3.4809\times10^{-5}
\eea
(\ref{F00}) is substantially smaller than (\ref{A00}), which is also seen   
by a direct comparison of (\ref{TWIST3XX}) to (\ref{AGTX}) 
\bea
\label{RATIO23}
\frac{\rm twist-2}{\rm twist-3}\approx 
\frac{(N_c-2)}{2N_c(N_c+2)} \frac 3{225}\frac {\delta q}{A_q}\approx  10^{-4}\nonumber\\
\eea
for $J_{II}/J_{IA}\approx 63/9=7$,
which is about the ratio of the gluonic charges (\ref{F00}) to (\ref{A00}). In the QCD instanton vacuum, the C-odd twist-3 contribution is substantially compared to the C-even twist-2 contribution, and vanishes in the
large $N_c$ limit. The extra suppression in $1/N_c$ stems
from the non-Abelian crossing and modular average (ringdots in Fig.~\ref{fig:inst_op}).

\subsection{Tripole form factors}
To proceed with the evaluation of the photoproduction of heavy pseudoscalar mesons,
we will use the tripole approximation for the gluonic form factors $A_g(t), C_g(t), f_g(t)$ with the charges $A_g(0), C_g(0), f_g(0)$ fixed by the QCD instanton vacuum. 
A full determination of the form factors, using the nucleon light front wavefunctions
derived in the QCD instanton vacuum in~\cite{Liu:2024}, with full account of the nucleon mechanical properties will be given elsewhere. With this in mind, the tripole form factors are
\bea
\label{ACF}
    A_g(t)&=&\frac{A_g(0)}{\left(1-\frac{t}{m^2_A}\right)^3}\nonumber\\
    C_g(t)&=&\frac{C_g(0)}{\left(1-\frac{t}{m^2_{C}}\right)^3}\nonumber\\
    f_g(t)&=&\frac{f_g(0)}{\left(1-\frac{t}{m^2_{3g}}\right)^3}
\eea
with the charges given by (\ref{A00}, \ref{C00}, \ref{F00}) respectively. 
The  tripole masses $m_A, m_{C}$ are fitted to the recent lattice data~\cite{Shanahan:2018pib}, with the results
\bea
m_{A}&=&2.02~\mathrm{GeV}\nonumber\\
m_{C}&=&1.226~\mathrm{GeV}\nonumber\\
m_{3g}&=&1.49~\mathrm{GeV}
\eea
The tripole mass $m_A$ reflects on the tensor  $2^{++}$ glueball mass in dual gravity, and the tripole mass $m_C$ on the mixing between the tensor  $2^{++}$ and scalar  $0^{++}$ glueball  masses. Similarly, The tripole mass $m_{3g}$ is fixed by dual gravity~\cite{Hechenberger:2024abg}, and reflects on the C-odd  $1^{+-}$ glueball mass.
Similar form factors were used using dual gravity~\cite{Mamo:2021krl,Hatta:2021can}
and the QCD factorization method~\cite{Guo:2021ibg,Guo:2023pqw}, in the analysis of coherent $J/\Psi$ production.

\begin{figure*}
\subfloat[\label{fig:dJPsi}]{%
\includegraphics[height=5cm,width=.45\linewidth]{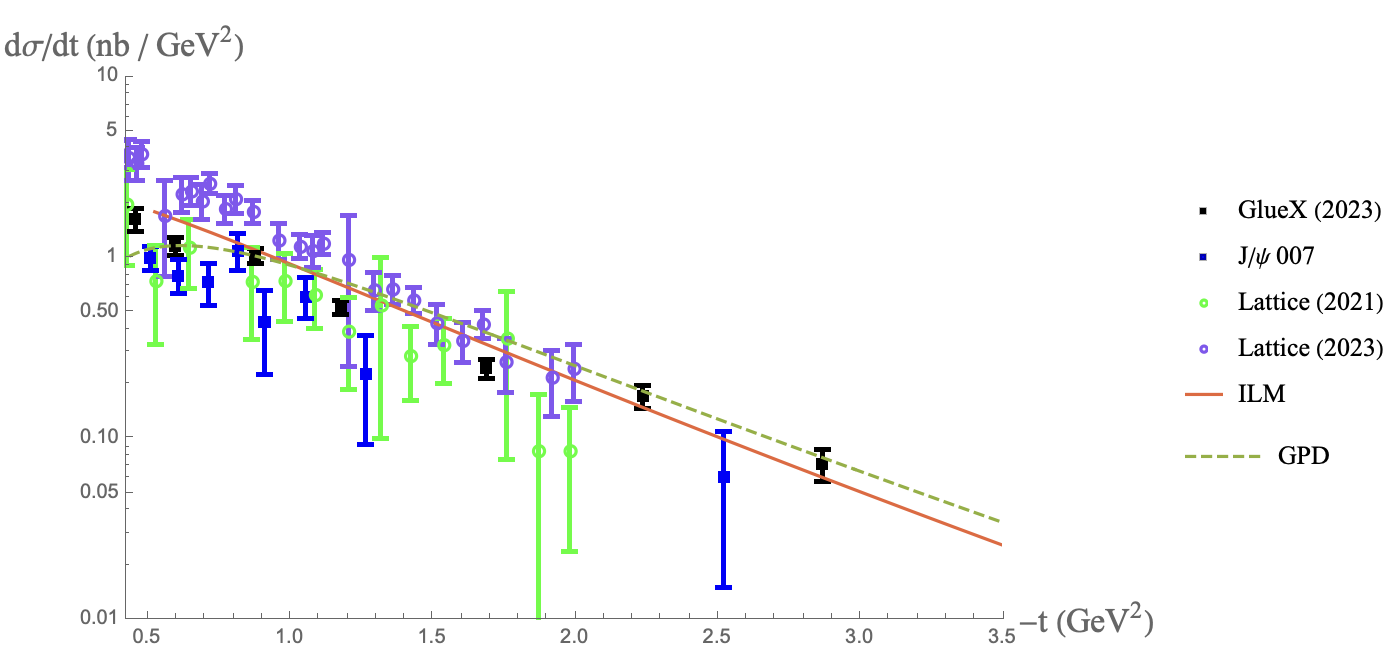}%
}\hfill
\subfloat[\label{fig:JPsi}]{%
\includegraphics[height=5cm,width=.45\linewidth]{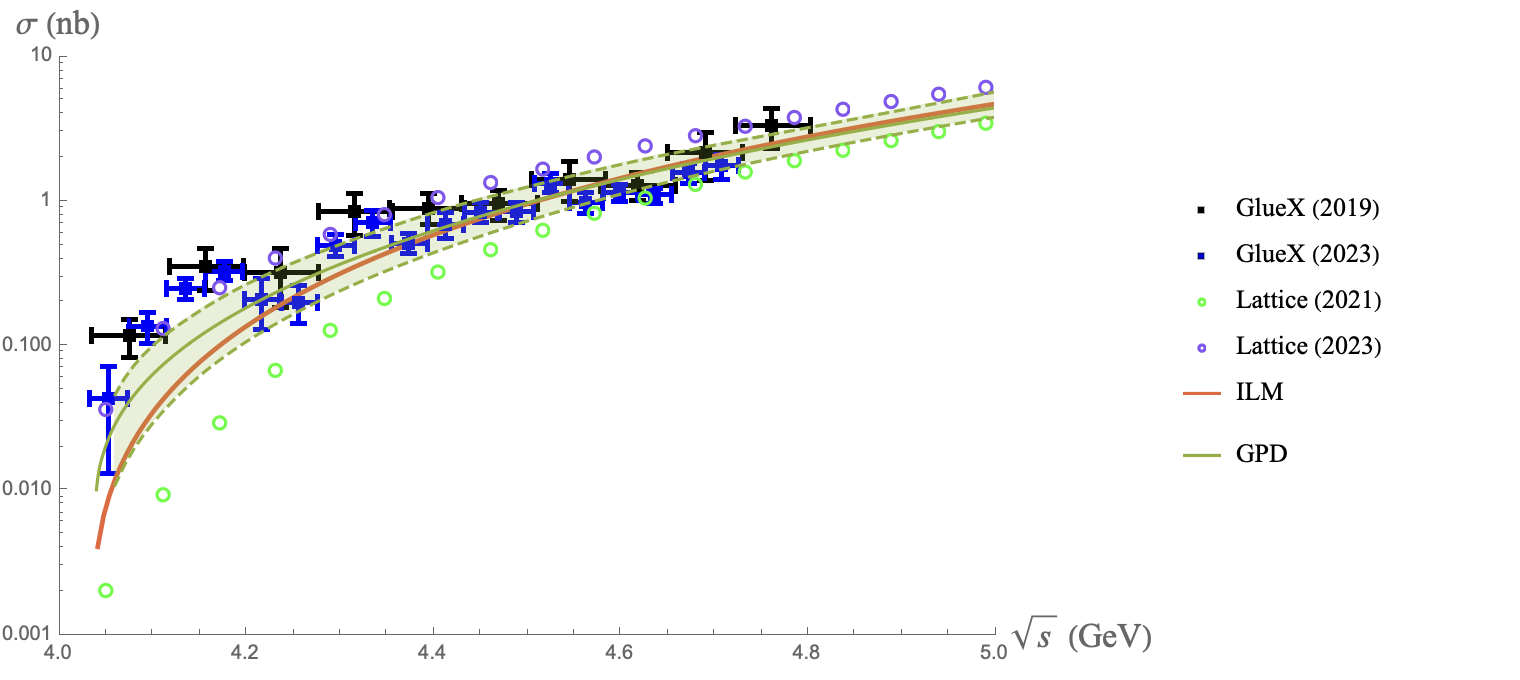}%
}\hfill
\subfloat[\label{fig:dUpsilon}]{%
\includegraphics[height=5cm,width=.45\linewidth]{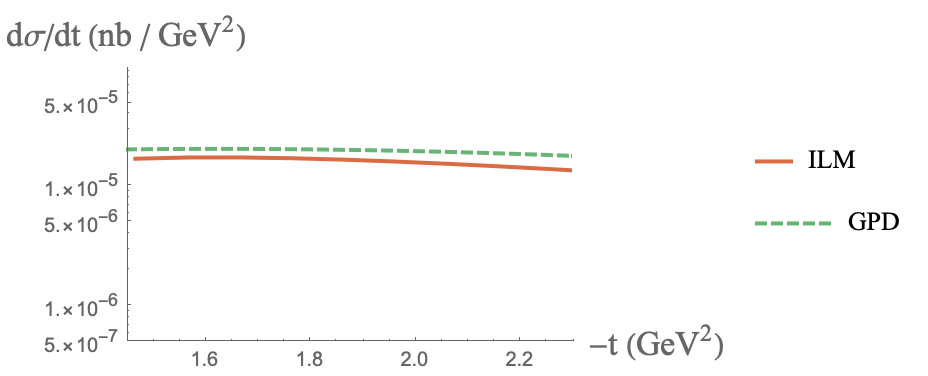}%
}\hfill
\subfloat[\label{fig:Upsilon}]{%
\includegraphics[height=5cm,width=.45\linewidth]{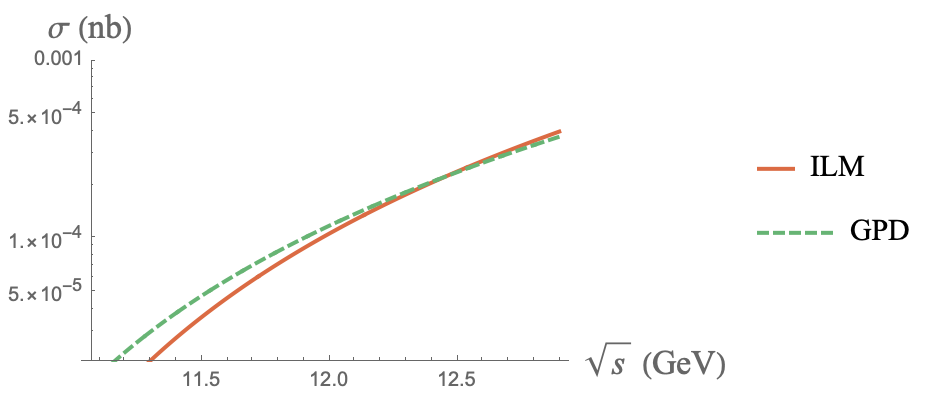}%
}
\caption{ a: The instanton estimation with the holographic prediction on the $t$-dependence \cite{Mamo:2022eui} using tripole approximation is compared to the GlueX experiment in 2023 \cite{GlueX:2023pev} and $J/\psi$ 007 \cite{Duran:2022xag}. The lattice calculations \cite{Pefkou:2021fni} and \cite{Hackett:2023rif} are also compared. We also compare with the GPD prediction \cite{Guo:2023pqw};
b: The instanton estimation with the holographic prediction on the $t$-dependence \cite{Mamo:2022eui} using tripole approximation is compared to the GlueX experiment in 2019 \cite{GlueX:2019mkq} and in 2023 \cite{GlueX:2023pev}. The lattice calculations \cite{Pefkou:2021fni} and \cite{Hackett:2023rif} are also compared. We also compare with the GPD prediction \cite{Guo:2023pqw};
c: The instanton estimation on the $\Upsilon$ production differential cross section; 
d: The cross section  with the holographic prediction on the $t$-dependence \cite{Mamo:2022eui} is compared to the GPD prediction \cite{Guo:2021ibg}.}
\end{figure*}

In Fig.~\ref{FFACF}a we show the gluonic gravitational form factor  $A_g(t)$ versus $-t=Q^2$ given in (\ref{ACF}) (orange-solid line) with the charge $A_g(0)$ fixed by the
QCD instanton vacuum parameters in (\ref{A00}). The same form factor used in the parameterization of the GPDs in~\cite{Guo:2023pqw} (green-dashed curve) is compared to the early lattice results~\cite{Pefkou:2021fni} with a large pion mass $m_\pi=450$ MeV 
(black data points), and the latest lattice results~\cite{Hackett:2023rif}
with a smaller pion mass $m_\pi=170$ MeV  
(blue data points). The gravitational charge from the QCD instanton vacuum is compatible
with the lattice results. The same comparison for the gluonic gravitational form factor $C_g(t)$ in (\ref{ACF}) is shown in Fig.~\ref{FFACF}b. Again, the value of the gravitational charge $C_g(0)$ from the QCD instanton vacuum is compatible with the
latest lattice results. The predicted C-odd form factor $f_g(t)$ in (\ref{ACF}) is
shown in Fig.~\ref{FFACF}c using the QCD instanton vacuum charge. This charge characterizes a C-odd gluonic twist-3 operator, which is strongly suppressed in the QCD instanton vacuum (see also below). The natural coupling to the pseudoparticles is through their charges $G\tilde G=\pm GG$, which explains the difference with the gluonic scalar and pseudoscalar form factors.

\section{Photoproduction cross sections}
\label{SEC4}
The differential cross sections for both the coherent production of heavy vector and pseudoscalar mesons off a nucleon, can now be evaluated in parallel for comparison,
and consistency with previous calculations for the vector mesons. 

\subsection{$J/\Psi, \Upsilon$ photo-production}
In the case of $J/\psi$ photoproduction, we average over the proton initial spin and sum over the final spin. The cross section reads
\begin{widetext}
\begin{equation}
\label{DIFFPSI}
\begin{aligned}
  \frac{d\sigma}{dt}=&\frac{Q_c^2e^2}{16\pi(s-M_N^2)^2}\sum_{\mathrm{polarizations}}|\mathcal{M}|^2  \\
=& 4\pi\alpha_{em}Q_c^2\frac{16\pi\alpha^2_s}{(s-M_N^2)^2}\frac{4}{N_cM^2_{J/\psi}}\left|\psi_{J/\psi}(0)\right|^2\frac{1}{4}\sum_{\lambda_\gamma \lambda_Xss'}(\epsilon_\gamma\cdot\epsilon_V^*)^2\left|\mathcal{W}_{2g}(t,\xi)\right|^2\\
=&4\pi \alpha_{em}Q_c^2\frac{16\pi\alpha^2_s}{(s-M_N^2)^2}\frac{4}{N_cM^2_{J/\psi}}\left|\psi_{J/\psi}(0)\right|^2\\
&\times\frac{4}{\xi^4}\left[(H_{2g}+E_{2g})^2(1-\xi^2)-2(H_{2g}+E_{2g})E_{2g}+\left(1-\frac{t}{4M_N^2}\right)E_{2g}^2\right]    
\end{aligned}
\end{equation}
\end{widetext}
The dominant amplitude for the photo-production of $J/\Psi$ stems solely from
their transverse polarizations~\cite{Sun:2021pyw}, where we used
$$\sum_{\lambda_\gamma \lambda_X}(\epsilon_\gamma\cdot\epsilon_V^*)^2=2$$
and in overall agreement with~\cite{Guo:2021ibg,Guo:2023pqw}. A comparison of 
\eqref{eq:f2g} with \eqref{eq:matrix_2g}, shows that the twist-$2$ functions $H_{2g}$ and $E_{2g}$ are related to the gluonic gravitational form factors $A_g,B_g,C_g$,
\bea
    H_{2g}(t,\xi)&=&A_g(t)+4\xi^2C_g(t)\nonumber\\
    E_{2g}(t,\xi)&=&B_g(t)-4\xi^2C_g(t)
\eea
The lattice simulations~\cite{Shanahan:2018pib} and dual gravity arguments~\cite{Liu:2024},
suggest that $B_g$ is about null. This will be assumed throughout. 

The heavy meson wave function is fixed by the decay constant (See Appendix.\ref{Appx:heavy_meson_WF}). For  $J/\psi$ ($\eta_c$ below), the strong coupling constant is fixed by the charmonium mass scale $\alpha_s(\mu=2m_c)=0.308$, with   
$M_{J/\psi}=3.097~\mathrm{GeV}$ 
($ M_{\eta_c}=2.984~\mathrm{GeV}$ below)~\cite{ParticleDataGroup:2018ovx}.
Similarly, for the heavier mesons  $\alpha_s(\mu=2m_b)=0.207$, with $M_{\Upsilon}=9.460~\mathrm{GeV}$ ($M_{\eta_b}=9.398~\mathrm{GeV}$ below)~\cite{ParticleDataGroup:2004fcd,ParticleDataGroup:2014cgo}.

In Fig.~\ref{fig:dJPsi} we show our result using the tripole 
form factors (\ref{ACF}) for the differential cross section for photo-production of $J/\Psi$  (\ref{DIFFPSI}) (orange-solid line),  in comparison to the  recently reported measurement by GlueX collaboration (black-data points) \cite{GlueX:2023pev} and by the $J/\Psi 007$ collaboration  (blue-data points) \cite{Duran:2022xag}. Also for comparison, we show the same  differential cross section using the lattice data for the form factors (green-open circles)
from~\cite{Liu:2024} and (purple-open circles) from~\cite{Liu:2024}, and from the  GPD analysis in~\cite{Guo:2023pqw}. The results for the total cross section for the same process and using the same labeling are shown in  Fig.~\ref{fig:JPsi}.

In Fig.~\ref{fig:dUpsilon} we show our prediction for the differential cross section for the coherent photoproduction of $\Upsilon$ (orange-solid line) 
in comparison to the GPD result (green-dashed line)~\cite{Guo:2023pqw}. In Fig.~\ref{fig:Upsilon} our prediction for the cross section for the same process (orange-solid line) also in comparison to the GPD result (green-dashed line)~\cite{Guo:2023pqw}.  The
results are totally compatible, since our approach follows their construction. In light of this, we now
proceed to the analysis of the coherent production of heavy pseudoscalars.

\begin{figure*}
\label{fig:eta_c_2}
\includegraphics[scale=0.5]{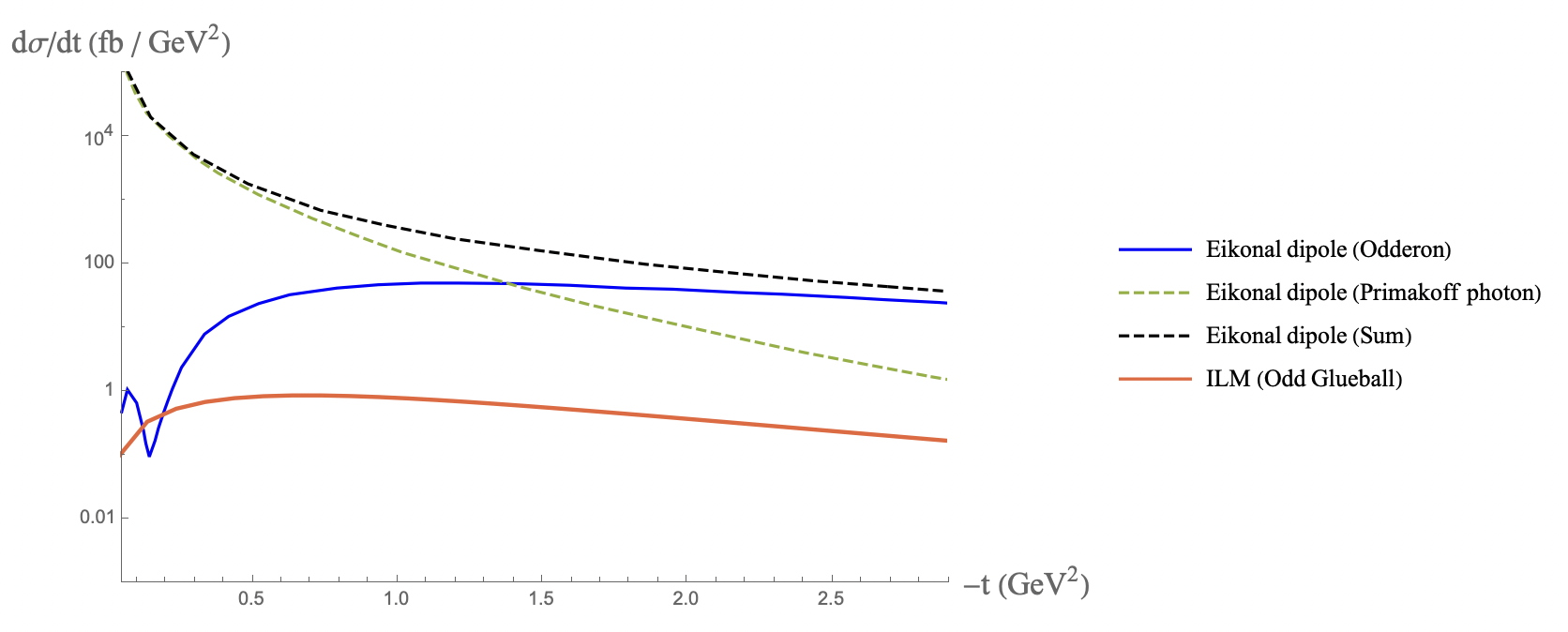}
\caption{ 
The instanton estimation on the $\eta_c$ production differential cross section with the holographic prediction on the $t$-dependence \cite{Hechenberger:2024abg} using fitted mass $m_{3g}=1.49~\mathrm{GeV}$ is compared to the model calculation using eikonal dipole approximation \cite{Dumitru:2019qec}.}
\end{figure*}

\begin{figure*}
\subfloat[\label{fig:eta_c}]{%
\includegraphics[height=5cm,width=.45\linewidth]{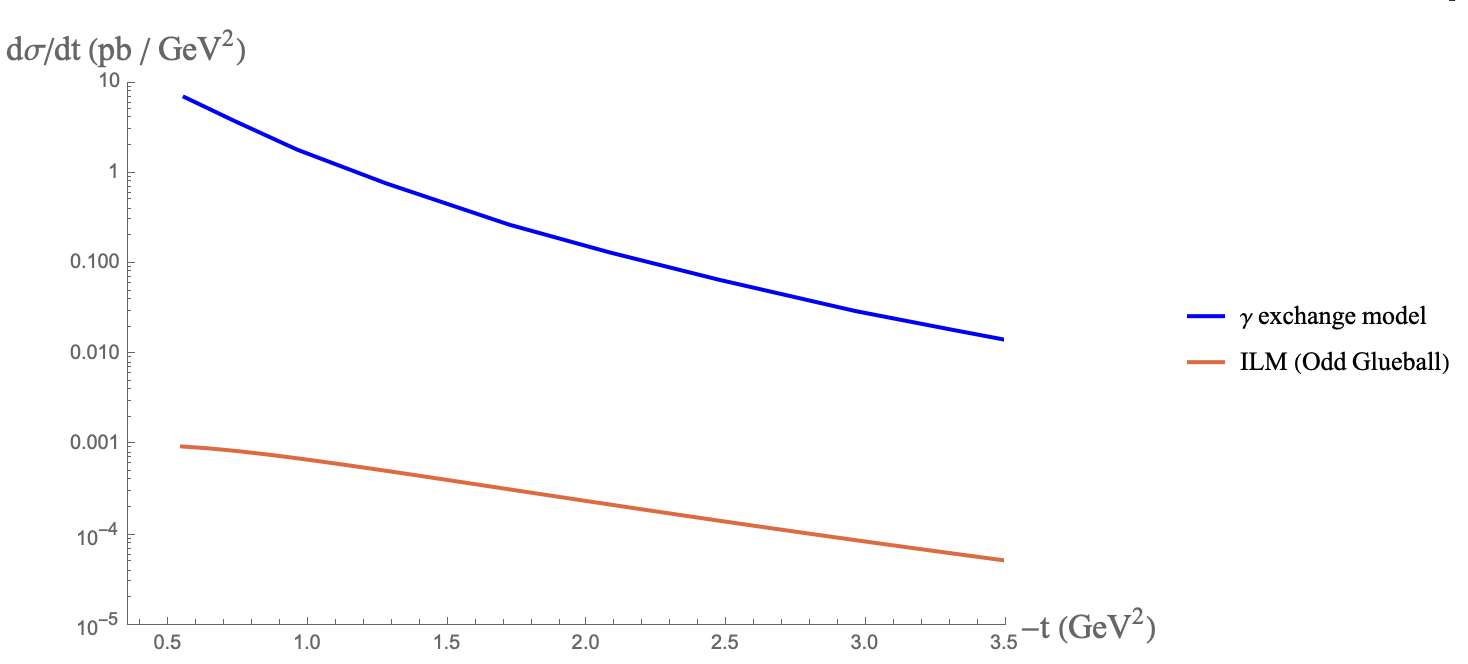}%
}\hfill
\subfloat[\label{fig:eta_b}]{%
\includegraphics[height=5cm,width=.45\linewidth]{dcs_eta_c.png}%
}
\caption{ The instanton estimation on the differential cross section of the $\eta_c$ production (a) and $\eta_b$ production (b) with the holographic prediction on the $t$-dependence \cite{Hechenberger:2024abg} using fitted mass $m_{3g}=1.49~\mathrm{GeV}$ is compared to the photon exchange process predicted in the perturbative QCD calculation \cite{Jia:2022oyl}.}
\end{figure*}

\subsection{$\eta_{c,b}$ photo-production}
The coherent photoproduction  heavy pseudoscalars  $\eta_{c,b}$ in the near threshold region, follows 
a similar reasoning. The corresponding differential cross section for $\eta_c$ production is given by
\begin{widetext}
\begin{equation}
\label{DTETAC}
\begin{aligned}
    \frac{d\sigma}{dt}=& 4\pi\alpha_{em}Q_c^2\frac{32\pi^2\alpha_s^3}{(s-M_N^2)^2}\frac{4}{N_cM^2_{\eta_c}}|\psi_{\eta_c}(0)|^2\frac{1}{4}\sum_{\lambda_\gamma ss'} \left|\epsilon_{\perp ij}\epsilon^i_{\gamma}\mathcal{W}^j_{3g}(t,\xi)\right|^2\\
    =&4\pi \alpha_{em}Q_c^2\frac{32\pi^2\alpha_s^3}{(s-M_N^2)^2}\frac{4}{N_cM^2_{\eta_c}}|\psi_{\eta_c}(0)|^2\left(\frac{3}{2}\right)^6\frac{36}{\xi^8}\left(\frac{M_N}{\bar{P}^+}\right)^2\\
    &\times\left[f_1^2(1-\xi^2)-2\xi f_1f_4-f_4^2\frac{t}{4M_N^2}+\frac{\Delta_\perp^2}{2M^2_N}L_1(t,\xi)-\frac{\bar{P}_\perp\cdot\Delta_\perp}{M^2_N}L_2(t,\xi)-\frac{\bar{P}_\perp^2}{2M^2_N}L_3(t,\xi)\right]
\end{aligned}
\end{equation}
The orbital form factor contributions are defined as
\begin{equation}
    L_1(t,\xi)=f_2^2\left(1-\frac{t}{4M^2_N}\right)+f^2_3(1-\xi^2)-f_1f_2+2f_2f_3-\xi f_3f_4-\frac{1}{2}f_4^2
\end{equation}
\begin{equation}
    L_2(t,\xi)=\xi f_1f_2+\frac{1}{2}f_1f_4+\frac{t}{4M^2_N}f_2f_4
\end{equation}
\begin{equation}
    L_3(t,\xi)=2\xi f_1f_4+\frac{t}{4M^2_N}f_4^2
\end{equation}
\end{widetext}
Using \eqref{eq:odd_gluball1}, we see that the twist-3 contributions  $f_{1,2,3,4}(t)$ are all tied to our C-odd form factor through
\bea
    f_1(t,\xi)&=&4\xi^2\frac{t}{4M^2_N}f_g(t)\nonumber\\
    f_2(t,\xi)&=&-f_3(t,\xi)=-4\xi^2f_g(t)\nonumber\\
    f_4(t,\xi)&=&8\xi^3f_g(t)
\eea

Note that in the forward limit $\Delta^\mu\rightarrow0$, the cross section vanishes as all twist-$3$ functions $f_{1,2,3,4}$ vanish in this limit.
Near the threshold region in the heavy meson limit, the allowed kinematic region is restricted to $\xi\rightarrow1$ and $t\rightarrow t_{th}$. The leftover of the center-of-mass energy $\sqrt{s}$ is not enough to excite a large orbital motion (large $\Delta_\perp$ and $\bar{P}_\perp$) during the scattering. Therefore, the momentum transfer $t$ would be constrained in a small range around $t_{th}\sim M_NM_X$. As the orbital motion inside the hadron bound state does not have significant effect in this regime, the remaining contribution to the differential cross section comes from the intrinsic property of the constituent quark which can be estimated in the instanton ensemble. The differential cross sections near the threshold in the heavy meson limit read

\begin{widetext}
\begin{equation}
    \frac{d\sigma}{dt}\bigg|_{\eta_c}\simeq4\pi \alpha_{em}Q_c^2\frac{32\pi^2\alpha_s^3}{(s-M_N^2)^2}\frac{4|\psi_{\eta_c}(0)|^2}{N_cM^2_{\eta_c}}\left(\frac{3}{2}\right)^6\frac{36}{\xi^8}\left(\frac{M_N}{\bar{P}^+}\right)^2\left[f_1^2(1-\xi^2)-2\xi f_1f_4-f_4^2\frac{t}{4M_N^2}\right]
\end{equation}
\end{widetext}

In Fig.~\ref{fig:eta_c_2} we show our result (\ref{fig:eta_c}) for the differential cross section for the coherent photo-production of $\eta_c$ near threshold (orange-solid line), to 
eikonalized odderon exchange (blue-solid curve), eikonalized photon exchange (green-dashed line) and their sum (black-dashed line)~\cite{Dumitru:2019qec}. Our 
result is substantially smaller than the one reported using the eikonal dipole approximation~\cite{Dumitru:2019qec}.
In Fig.~\ref{fig:eta_c} we compare our result for the $\eta_c$ production (orange-solid line), to the photon exchange model (blue-solid line)~\cite{Jia:2022oyl}. The same comparison is
carried in Fig.~\ref{fig:eta_b} for the $\eta_b$ production.  The photon rate dwarfs our estimate for the  C-odd gluon exchange production of heavy $\eta_{c,b}$ near threshold.

In the threshold region and in the heavy meson limit ($\xi\rightarrow1$, $t\rightarrow t_{th}$), the ratios of the vector to pseudoscalar differential cross sections for the charmed and bottom mesons respectively, are
\bea
    \frac{d\sigma/dt({\eta_c})}{d\sigma/dt({J/\psi})}\Bigg|_{\substack{\sqrt{s}=4.05~ \mathrm{GeV} \\ -t=1.8~\mathrm{GeV}^2}}&=&9.782\times10^{-6}\nonumber\\
     \frac{d\sigma/dt({\eta_b})}{d\sigma/dt({\Upsilon})}\Bigg|_{\substack{\sqrt{s}=10.4~ \mathrm{GeV} \\ -t=7.2~\mathrm{GeV}^2}}&=&1.510\times10^{-6}\nonumber\\
\eea
This illustrates the smallness of the gluonic mechanism for pseudoscalar production, using QCD factorisation.  These ratios depend sensitively on the C-even twist-2 gluonic gravitational charges $A_g$, and the C-odd twist-3 gluonic charge, as shown  in (\ref{RATIO23}).

Finally, we note that in dual gravity the photo-production of heavy pseudoscalars in the threshold region, was recently found to be substantially larger~\cite{Hechenberger:2024abg}. This is due to the fact that the boundary dual of the bulk Kalb-Ramond field, is a twist-5 operator, with a matrix element in the QCD instanton vacuum~\cite{Liu:2024}
\begin{widetext}
\bea
\label{TWIST5}
    \langle P'|d^{abc}F^{a}_{\mu\nu}F^{b}_{\rho\lambda}F^{c}_{\rho\lambda}|P\rangle\approx -\frac{N_c-2}{2N_c^2(N_c^2-1)}\left(\frac{n_{I+A}}{2}\frac{4\pi^2\rho^2}{m^*}\right)^2\frac{16\pi^2}{9m^*}q^2\langle P'|\bar{\psi}\sigma_{\mu\nu}\psi|P\rangle
\eea
\end{widetext}
In contrast, the QCD factorization process is driven by the twist-3 operator, with the matrix element in the QCD instanton vacuum (\ref{TWIST3X}). A comparison of (\ref{TWIST5}) to
(\ref{TWIST3X}) shows that 
\bea
\frac{\rm twist-3}{\rm twist-5}\approx 
\bigg(\frac 1{N_c+2}\frac 98\frac 1{225}\bigg)\,(\rho q)^2\approx 4\,10^{-4}\nonumber\\
\eea
with the twist-3 parametrically suppressed near threshold $\rho q\approx 1$, and again vanishingly small in the large $N_c$ limit.

\section{Conclusions}
\label{SEC5}
Threshold coherent photo-production of heavy mesons at current and future electron facilities, has the potential of probing the gluonic content of the nucleon at low resolution through pertinent C-even and C-odd
gluonic form factors, as shown by the recent experiments carried at JLAB~\cite{GlueX:2019mkq,Duran:2022xag,GlueX:2023pev}. 
The C-even nucleon gluonic gravitational  form factors follows from the production  of heavy $J/\Psi,\Upsilon$. Their analysis was carried by many, with most notably the QCD factorization method~\cite{Guo:2021aik,Guo:2023pqw} and the gravity dual method~\cite{Mamo:2019mka,Mamo:2022eui} (and references therein). 

The C-odd nucleon gluonic form factors follow from the production 
of heavy $\eta_{c,b}$. Both the C-even and C-odd form factors depend sensitively on their gluonic charges at $t=0$. While the formers are presently accessible to 
QCD lattice simulations~\cite{Hackett:2023rif}, the latters are not. To this effect, we have used the QCD instanton vacuum to 
evaluate these charges, with a good agreement with the 
lattice results for the C-odd charges.

We have analysed the photo-production of heavy vector and pseudoscalar mesons near the threshold region, following the  QCD factorization method~\cite{Guo:2021aik,Guo:2023pqw}, using our newly
estimated gluonic charges from the QCD instanton vacuum. 
The results for $J/\Psi, \Upsilon$ production are in good agreement with those estimated in~\cite{Mamo:2022eui,Guo:2023pqw}, and the recent JLAB measurements~\cite{Duran:2022xag}. 
The results for threshold differential cross sections for photo-production  of $\eta_{c,b}$ are substantially smaller than those of  $J/\Psi, \Upsilon$, and dwarfed by  the C-odd exchange of a single photon.

In the QCD factorization method, coherent heavy vector meson production near threshold, is driven by a leading C-even twist-2 gluonic operator at large skewness, in comparison to 
heavy pseudoscalar meson production, which is driven by by a C-odd twist-3 gluonic operator at large skewness. The latter operator is parametrically suppressed in the QCD instanton vacuum.

The threshold photo-production cross sections for heavy mesons are very sensitive to the gluonic charges. In the
QCD vacuum these charges are penalized by the pseudoparticles density squared, with the C-odd charges substantially more than the C-even charges, by further factors of $1/N_c$ from the modular integrations. This makes  the measurement of this gluonic charge and form factor, the more pertinent for a QCD lattice simulation.

The present production mechanism  for photoproduction of heavy pseudoscalars, is based on QCD factorization. It is driven by a C-odd twist-3 gluonic operator which is seen to be substantially suppressed in the threshold region. In contrast,  dual gravity at strong coupling shows that the same production mechanism is driven by 
a C-odd twist-5 gluonic operator, which is substantially
larger in the threshold region \cite{Hechenberger:2024abg}.

\begin{acknowledgments}
This work is supported by the Office of Science, U.S. Department of Energy under Contract No. DE-FG-88ER40388. This research is also supported in part within the framework of the Quark-Gluon Tomography (QGT) Topical Collaboration, under contract no. DE-SC0023646.
\end{acknowledgments}

\appendix

\section{Gluonic form factors in the QCD instanton vacuum}
\label{SEC3X}
The C-even and C-odd gluonic matrix elements in the proton state at low momentum transfer, can be estimated in the QCD instanton vacuum. We will give a brief presentation of the method  below, with more details to be presented in~\cite{Liu:2024}.
More specifically, a general transition matrix element of a gluon operator $\mathcal{O}[A]$ in a fixed in-out hadron  state, is given by the grand-cannonical average~\cite{Liu:2024}
(and references therein)
\begin{widetext}
\begin{equation}
\begin{aligned}
\label{eq:o_had_exp}
    &\langle P'| \mathcal{O}[A]|P\rangle_{N_\pm}=\sum_{n=1}^\infty\frac{1}{n!}\left[\sum_{k=0}^n\binom{n}{k}\left(\frac{N_+}{V}\right)^{n-k}\left(\frac{N_-}{V}\right)^k\left(\prod_{f}\frac{4\pi^2\rho^2}{m^*}\right)^{n}\langle P'|\mathcal{O}_{++\cdots-}|P\rangle_{\mathrm{eff}}\right]\\
    =&\frac{N_+}{V}\left(\frac{4\pi^2\rho^2}{m^*}\right)^{N_f}\langle P'|\mathcal{O}_+|P\rangle_{\mathrm{eff}}+\frac{N_-}{V}\left(\frac{4\pi^2\rho^2}{m^*}\right)^{N_f}\langle P'|\mathcal{O}_-|P\rangle_{\mathrm{eff}}+\frac{1}{2}\frac{N^2_+}{V^2}\left(\frac{4\pi^2\rho^2}{m^*}\right)^{2N_f}\langle P'|\mathcal{O}_{++}|P\rangle_{\mathrm{eff}}\\
    &+\frac{N_+N_-}{V^2}\left(\frac{4\pi^2\rho^2}{m^*}\right)^{2N_f}\langle P'|\mathcal{O}_{+-}|P\rangle_{\mathrm{eff}}+\frac{1}{2}\frac{N^2_-}{V^2}\left(\frac{4\pi^2\rho^2}{m^*}\right)^{2N_f}\langle P'|\mathcal{O}_{--}|P\rangle_{\mathrm{eff}}+\cdots
\end{aligned}
\end{equation}
\end{widetext}
with the sum running over the number of $N_\pm$ instantons and anti-instantons (pseudoparticles). The labeling in the gluonic entries $\mathcal{O}_{++\cdots-}$ refers to the fixed $N_\pm$ in the sum. Each of these matrix elements is evaluated in an effective field theory over the pseudoparticles moduli (modular gluons and quarks), given by
\begin{widetext}
\begin{equation}
\begin{aligned}
\label{eq:effective_action}
    \mathcal{L}_{\mathrm{eff}}(N_+,N_-)=&\left[-\psi^\dagger (i\slashed{\partial}-m^*_f)\psi+\frac{1}{4}(G^a_{\mu\nu})^2\right]
    -G\theta_+-G\theta_-
\end{aligned}
\end{equation}
For a dilute ensemble, each emerging 't Hooft vertex $\theta_{\pm}$ follows from the random averaging over the single pseudoparticle moduli with mean size fixed,
\begin{align}
\label{eq:tHooft}
\theta_{+}=&\int dU_{I}\prod_f\left[\frac{m_f}{4\pi^2\rho^2}+i\psi^\dagger_f(x)U_I\frac{1}{2}\left(1+\frac{1}{4}\tau^a\bar{\eta}^a_{\mu\nu}\sigma^{\mu\nu}\right)U_I^\dagger\frac{1-\gamma^5}{2}\psi_f(x)\right]e^{-\frac{2\pi^2}{g}\rho^2R^{ab}(U_I)\bar{\eta}^b_{\mu\nu}G^a_{\mu\nu}}\nonumber\\
\theta_{-}=&\int dU_{A}\prod_f\left[\frac{m_f}{4\pi^2\rho^2}+i\psi^\dagger_f(x)U_A\frac{1}{2}\left(1+\frac{1}{4}\tau^a\eta^a_{\mu\nu}\sigma^{\mu\nu}\right)U_A^\dagger\frac{1+\gamma^5}{2}\psi_f(x)\right]e^{-\frac{2\pi^2}{g}\rho^2R^{ab}(U_A)\eta^b_{\mu\nu}G^a_{\mu\nu}} 
\end{align}
\end{widetext}

The matrix elements are evaluated  order by order in
the pseudoparticle density. The emergent 't Hooft coupling $G$,  is fixed by the saddle point approximation,
\begin{equation}
    G=\frac{N_++N_-}{V}
    \frac{(4\pi^2\rho^3)^{N_f}}{\prod_f(\rho m^*_f)}
\end{equation}
for a fixed  mean instanton size $\rho$, pseudoparticle density $N/V$,  and   determinantal mass $m^*_f$~\cite{Schafer:1995pz,Faccioli:2001ug,Shuryak:2021fsu}
\begin{equation}
\label{eq:m_det}
    m_f^*=m_f-\frac{2\pi^2\rho^2}{N_c}\langle\bar{\psi}_f\psi_f\rangle
\end{equation}
The vacuum parameters are fixed to $\rho=0.31~\mathrm{fm}$ and $n_{I+A}=1~\mathrm{fm}^{-4}$, and a dilute packing fraction
\bea
\label{PACKING}
\kappa=n_{I+A}\pi^2\rho^4\approx 0.1
\eea
For a quark current mass $m=(m_u+m_d)/2\simeq 6.0$ MeV, the determinantal mass $m^*$ is given by \eqref{eq:m_det}
$$
m^*=159.92 ~\mathrm{MeV}
$$
which is to be compared to the heavier constituent 
quark mass $M(0)\approx 395\,{\rm MeV}$. The determinantal
mass limits the hopping of the modular quark zero modes to the nearest neighbor  pseudoparticles (SIA). Both masses are close to those used in~\cite{Faccioli:2001ug,Shuryak:2021fsu,Liu:2023yuj,Liu:2023fpj}.
The quark condensate $\langle \bar{\psi}\psi\rangle$ is given by the resummation of the nearest quark hopping (\ref{eq:qq}), numerically close to the one given in \cite{Ioffe:2002ee}
\begin{equation}
\label{eq:qq}
\langle \bar{\psi}\psi\rangle=-\frac{n_{I+A}}{m^*}\simeq-(211.6~\mathrm{MeV})^3
\end{equation}
For completeness, we refer to~\cite{Diakonov:1995qy, Schafer:1996wv}, for more details regarding the phenomenology of the QCD instanton vacuum. 

The form factors following from (\ref{eq:o_had_exp}) can be expanded systematically, in terms of the instanton density. 
Translational symmetry  relates the hadronic matrix element of $\mathcal{O}[A]$ to the momentum transfer  between the hadronic states, 
\begin{align}
\label{OFFMAT}
     \langle P'|\mathcal{O}[A]|P\rangle = \frac{1}{V}\int d^4x\langle P'|\mathcal{O}[A(x)]|P\rangle e^{-i\Delta\cdot x}\nonumber\\
\end{align}
The recoiling hadron momentum is defined as $P'=P+\Delta$, and the forward limit follows from $\Delta\rightarrow0$.
(\ref{eq:o_had_exp}) generalizes the arguments in~\cite{Weiss:2021kpt} to off-forward with pseudoparticle
correlations. The latters dominate a number of light front 
matrix elements.

\begin{figure*}
\hfill
\subfloat[\label{fig:inst_op_2}]{%
    \includegraphics[scale=0.8]{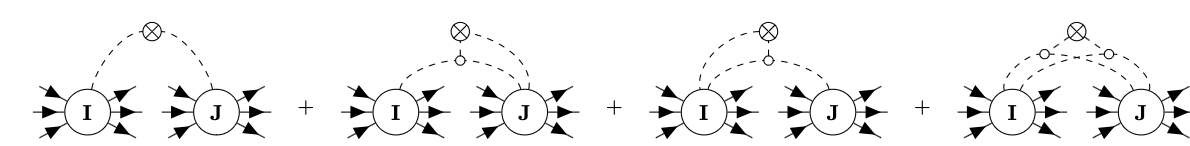}}
\hfill
\subfloat[\label{fig:inst_op_3}]{%
 \includegraphics[scale=0.8]{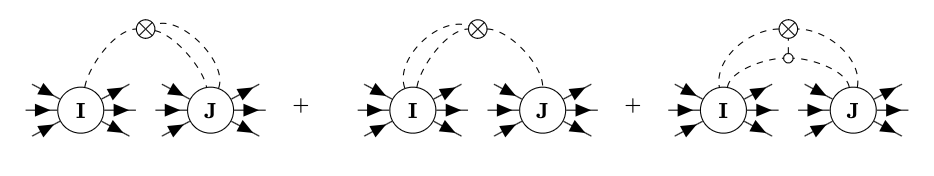}}
\caption{The diagrams of $\mathcal{O}[A_I,A_J]$ in the multi-instanton expansion of the two-gluon operator $\mathcal{O}_{2g}$. Each line connected to the instantons $I$ and $J$ represents the gluon fields in the operator. Each of the ringdots in the diagrams represent the insertion of the non-Abelian cross term $G_{\mu\nu}[A_I,A_J]$. 
Mass decomposition using Ji's nucleon mass sum rule, in the QCD instanton vacuum at the resolution $\mu=636~\mathrm{MeV}
\sim1/\rho$ (a), and after DGLAP evolution at a resolution 
$\mu=2~\mathrm{GeV}$ (b).
}
\label{fig:inst_op}
\end{figure*}

The evaluation of (\ref{OFFMAT}) for a 2-gluon operator sourced by the pseudoparticles IJ, is illustrated in Fig.\ref{fig:inst_op_2}.
The dashed lines refer to the modular gluons. More specifically, since each of the external fermion flavor $N_f$ contributes a pair of modular gluons $UU^\dagger$ in the color group integral, each
averaging yields a factor of $1/N_c^{N_f}$. 

While considering the contribution of clusters of pseuparticles, it helps to note that 
the dominant contribution stems from closed pairs, where we may 
approximate  (R-expansion)
\bea
\label{RAPP}\bar{\psi}(z_I)\psi(z_J)\simeq\bar{\psi}(z)\psi(z)-R_\mu\bar{\psi}(z)\overleftrightarrow{\partial_\mu}\psi(z)\nonumber\\
\eea
with  $R=z_I-z_J$ the relative distance. With this in mind, 
the nearest quark hopping within a pseudoparticle  $J_{II}(\rho m^*)$ is
\begin{equation}
\begin{aligned}
\label{eq:quark_hopping_II}
&J_{II}(\rho m^*)=\left(\frac{1}{\rho m^*}\right)^3\approx63.04
\end{aligned}
\end{equation}
and between pseudoparticles  $J_{IA}(\rho m^*)$ is 
\begin{widetext}
\begin{equation}
\begin{aligned}
\label{eq:quark_hopping_IA}
J_{IA}(\rho m^*)
=\frac{1}{16\pi^2}\left(\frac{1}{\rho m^*}\right)^2\int d^4R\int_0^\infty dk\frac{k^3\mathcal{F}(k)J_2(kR)}{k^2+\rho^2(m^*)^2}
\approx&~8.766
\end{aligned}
\end{equation}
\end{widetext}
where we used the nearest-neighbor quark propagator with
a determinantal mass (SIA)
\begin{equation}
    S(x-y)\sim \int \frac{d^4k}{(2\pi)^4}\frac{i\slashed{k}+m^*}{k^2+m^{*2}}\mathcal{F}(\rho k)e^{-ik\cdot(x-y)}
\end{equation}

\section{Light front wave function for heavy mesons}
\label{Appx:heavy_meson_WF}
Generally, the light front wave function for mesons are defined as
\begin{widetext}
\begin{equation}
    |X(P)\rangle=\int\frac{dk^+d^2k_\perp}{(2\pi)^3}\Psi_X(k,s_1,s_2)b^\dagger_{s_1}(k)c^\dagger_{s_2}(P-k)|0\rangle
\end{equation}
For vector quarkonium, the light front wave function is defined as \cite{Liu:2023fpj}
\begin{equation}
    \Psi_X(k,s_1,s_2)=\frac{1}{\sqrt{N_c}}\frac{\phi_{X}(k)}{\sqrt{2k^+(M_Xv-k)^{+}}}\bar{u}_{s_1}(k_1)\gamma\cdot\epsilon_{V} v_{s_2}(k_2)
\end{equation}
and for pseudoscalar quarkonium, the light front wave function is defined as \cite{Liu:2023yuj}
\begin{equation}
    \Psi_X(k,s_1,s_2)=\frac{1}{\sqrt{N_c}}\frac{\phi_{X}(k)}{\sqrt{2k^+(M_Xv-k)^{+}}}\bar{u}_{s_1}(k_1)i\gamma^5 v_{s_2}(k_2)
\end{equation}
\end{widetext}
where $v^\mu$ is the four velocity of the quarkonium and $\phi_X(k)$ is the spin-independent wave function in momentum space.

In heavy limit, the relative momentum between quark and antiquark is of order $\mathcal{O}(\alpha_sM_X)$. Thus at the leading order of $\alpha_s$, the wave function can be further simplied by

\begin{equation}
    \phi_{X}(k)\simeq (2\pi)^3\delta(k^+-M_Xv^+/2)\delta^2(k_\perp)\psi_X(0)
\end{equation}
where we already assume $M_X=2m_Q$ for the quarkonium at the leading order of $\alpha_s$ and $\psi_X$ is the heavy meson wave function  
$$
\psi_X(0)=\int\frac{dk^+d^2k_\perp}{(2\pi)^3}\phi_X(k)
$$

Thus, in the heavy limit, the wave function for $J/\psi$ and the wave function for $\eta_c$ reads

\begin{widetext}
\begin{equation}
\begin{aligned}
    |J/\psi\rangle=\frac{1}{\sqrt{N_c}}\frac{\psi_{J/\psi}(0)}{\sqrt{2}m_cv^+}\bar{u}_{s_1}(m_cv)\gamma\cdot\epsilon_{V} v_{s_2}(m_cv)b^\dagger_{s_1}(m_cv)c^\dagger_{s_2}(m_cv)|0\rangle
\end{aligned}
\end{equation}

\begin{equation}
\begin{aligned}
    |\eta_c\rangle=\frac{1}{\sqrt{N_c}}\frac{\psi_{\eta_c}(0)}{\sqrt{2}m_cv^+}\bar{u}_{s_1}(m_cv)i\gamma^5v_{s_2}(m_cv)b^\dagger_{s_1}(m_cv)c^\dagger_{s_2}(m_cv)|0\rangle
\end{aligned}
\end{equation}
\end{widetext}
The heavy wave function can be estimated by the decay rate with the NLO QCD radiative correction included.

For heavy vector mesons \cite{Bodwin:2006yd}
\begin{widetext}
\begin{equation}
    \Gamma(J/\psi\rightarrow e^+e^-)=\frac{64\pi\alpha_{em}^2Q_c^2}{3M_{J/\psi}}N_c|\psi_{J/\psi}(0)|^2\left(1-\frac{16}{3}\frac{\alpha_s}{\pi}\right)
\end{equation}
For heavy pseudoscalar mesons \cite{Lansberg:2008cq,Fabiano:2002se}:
\begin{equation}
    \Gamma(\eta_c\rightarrow \gamma\gamma)=\frac{64\pi\alpha_{em}^2Q_c^4}{M_{\eta_c}}N_c|\psi_{\eta_c}(0)|^2\left(1-\frac{20-\pi^2}{3}\frac{\alpha_s}{\pi}\right)
\end{equation}
\end{widetext}
Note that our non-relativistic wavefunctions maps  onto 
those in~\cite{Guo:2021aik} through $$4M_V|\psi_V|^2\rightarrow |\psi_V|^2$$
The experimental decay rates for each heavy meson are shown in Table \ref{tab:decay_rate}. Each value is obtained by PDG group in \cite{ParticleDataGroup:2018ovx}.
With the QED coupling $\alpha_{em}$ at charmonium mass scale fixed to be $1/134$ and at bottomonia mass scale $1/132$ \cite{Erler:1998sy}, thus, we have
\bea
    |\psi_{J/\psi}(0)|^2&=&7.237 \times10^{-3}\nonumber\\
     |\psi_{\eta_c}(0)|^2&=& 3.361\times10^{-3}\nonumber\\
    |\psi_{\Upsilon}(0)|^2&=&1.463\times10^{-2}\nonumber\\
    |\psi_{\eta_b}(0)|^2&=& 1.463\times10^{-2}
\eea

\begin{table}
    \centering
    \begin{tabular}{c|c}
         Decay modes    & (keV) \\
    \hline
      $\Gamma(J/\psi\rightarrow e^+e^-)$ & 5.55\\ $\Gamma(\eta_c\rightarrow \gamma\gamma)$  & 5.0 \\
      $\Gamma(\Upsilon\rightarrow e^+e^-)$ & 1.29\\
      $\Gamma(\eta_b\rightarrow \gamma\gamma)$ & 0.52
    \end{tabular}
    \caption{The decay rates of the heavy vector meson decay to electron-positron pair and the heavy pseudoscalar meson decay to two photons are listed. $\eta_b$ decay is not available from the experiment. The value is estimated by assuming $|\psi_{\Upsilon}(0)|^2=|\psi_{\eta_b}(0)|^2$ based on the heavy quark symmetry.}
    \label{tab:decay_rate}
\end{table}

\section{Glueball excitations in twist expansion}
\label{Appx:OPE}
The operator product expansion of the nulti-gluon distributions are shown in this Appendix. The glueball states in the Regge trajectory of the gluon exchange process are probed by each of the local operators inside the hadronic bound states.

\begin{widetext}

In the process of two gluon exchange, the twist expansion corresponds to the Regge trajectory with the lowest $2^{++}$ glueball. 
\begin{equation}
\begin{aligned}
F^{a+i}\left(-\lambda^-/2\right) F^{a+}{}_i\left(\lambda^-/2\right)=&\sum_{n_1,n_2=0}^\infty\left(i\frac{\lambda^-}{2}\right)^{n_1}\left(-i\frac{\lambda^-}{2}\right)^{n_1}\frac{(iD^+)^{n_1}}{n_1!}F^{a+i}\frac{(iD^+)^{n_2}}{n_2!}F^{a+}{}_i
\end{aligned}
\end{equation}

In the process of two gluon exchange, the twist expansion corresponds to the Regge trajectory with the lowest $1^{+-}$ glueball.
 
\bea
    &&d^{abc}\tilde{F}^{a+\mu}\left(-\frac{2}{3}\rho^-\right) F^{b+i}\left(-\frac{1}{2}\lambda^-+\frac{1}{3}\rho^-\right)F^{c+}{}_i\left(\frac{1}{2}\lambda^-+\frac{1}{3}\rho^-\right)\nonumber\\
    =&&d^{abc}\sum_{n_1,n_2,n_3=0}^{\infty}\left(-\frac{2}{3}\rho^-\right)^{n_1}\left(-\frac{1}{2}\lambda^-+\frac{1}{3}\rho^-\right)^{n_2}\left(\frac{1}{2}\lambda^-+\frac{1}{3}\rho^-\right)^{n_3}\nonumber\\
    &&\qquad\qquad\qquad\times\frac{(iD^+)^{n_1}}{n_1!}\tilde{F}^{a+\mu}\left(0\right)\frac{(iD^+)^{n_2}}{n_2!} F^{b+i}\left(0\right)\frac{(iD^+)^{n_3}}{n_3!}F^{c+}{}_i\left(0\right)
\eea
\end{widetext}
Among these operators with the same twist, operators with higher dimension corresponds to higher glueball excitations in the Regge trajectory.

\bibliography{ref}

\begin{thebibliography}{61}%
\makeatletter
\providecommand \@ifxundefined [1]{%
 \@ifx{#1\undefined}
}%
\providecommand \@ifnum [1]{%
 \ifnum #1\expandafter \@firstoftwo
 \else \expandafter \@secondoftwo
 \fi
}%
\providecommand \@ifx [1]{%
 \ifx #1\expandafter \@firstoftwo
 \else \expandafter \@secondoftwo
 \fi
}%
\providecommand \natexlab [1]{#1}%
\providecommand \enquote  [1]{``#1''}%
\providecommand \bibnamefont  [1]{#1}%
\providecommand \bibfnamefont [1]{#1}%
\providecommand \citenamefont [1]{#1}%
\providecommand \href@noop [0]{\@secondoftwo}%
\providecommand \href [0]{\begingroup \@sanitize@url \@href}%
\providecommand \@href[1]{\@@startlink{#1}\@@href}%
\providecommand \@@href[1]{\endgroup#1\@@endlink}%
\providecommand \@sanitize@url [0]{\catcode `\\12\catcode `\$12\catcode
  `\&12\catcode `\#12\catcode `\^12\catcode `\_12\catcode `\%12\relax}%
\providecommand \@@startlink[1]{}%
\providecommand \@@endlink[0]{}%
\providecommand \url  [0]{\begingroup\@sanitize@url \@url }%
\providecommand \@url [1]{\endgroup\@href {#1}{\urlprefix }}%
\providecommand \urlprefix  [0]{URL }%
\providecommand \Eprint [0]{\href }%
\providecommand \doibase [0]{http://dx.doi.org/}%
\providecommand \selectlanguage [0]{\@gobble}%
\providecommand \bibinfo  [0]{\@secondoftwo}%
\providecommand \bibfield  [0]{\@secondoftwo}%
\providecommand \translation [1]{[#1]}%
\providecommand \BibitemOpen [0]{}%
\providecommand \bibitemStop [0]{}%
\providecommand \bibitemNoStop [0]{.\EOS\space}%
\providecommand \EOS [0]{\spacefactor3000\relax}%
\providecommand \BibitemShut  [1]{\csname bibitem#1\endcsname}%
\let\auto@bib@innerbib\@empty
\bibitem [{\citenamefont {Sch\"afer}\ and\ \citenamefont
  {Shuryak}(1998)}]{Schafer:1996wv}%
  \BibitemOpen
  \bibfield  {author} {\bibinfo {author} {\bibfnamefont {Thomas}\ \bibnamefont
  {Sch\"afer}}\ and\ \bibinfo {author} {\bibfnamefont {Edward~V.}\ \bibnamefont
  {Shuryak}},\ }\bibfield  {title} {\enquote {\bibinfo {title} {{Instantons in
  QCD}},}\ }\href {\doibase 10.1103/RevModPhys.70.323} {\bibfield  {journal}
  {\bibinfo  {journal} {Rev. Mod. Phys.}\ }\textbf {\bibinfo {volume} {70}},\
  \bibinfo {pages} {323--426} (\bibinfo {year} {1998})},\ \Eprint
  {http://arxiv.org/abs/hep-ph/9610451} {arXiv:hep-ph/9610451} \BibitemShut
  {NoStop}%
\bibitem [{\citenamefont {Shuryak}(2018)}]{Shuryak:2018fjr}%
  \BibitemOpen
  \bibfield  {author} {\bibinfo {author} {\bibfnamefont {Edward}\ \bibnamefont
  {Shuryak}},\ }\bibfield  {title} {\enquote {\bibinfo {title} {{Lectures on
  nonperturbative QCD ( Nonperturbative Topological Phenomena in QCD and
  Related Theories)}},}\ }\href@noop {} {\  (\bibinfo {year} {2018})},\ \Eprint
  {http://arxiv.org/abs/1812.01509} {arXiv:1812.01509 [hep-ph]} \BibitemShut
  {NoStop}%
\bibitem [{\citenamefont {Zahed}(2021)}]{Zahed:2021fxk}%
  \BibitemOpen
  \bibfield  {author} {\bibinfo {author} {\bibfnamefont {Ismail}\ \bibnamefont
  {Zahed}},\ }\bibfield  {title} {\enquote {\bibinfo {title} {{Mass sum rule of
  hadrons in the QCD instanton vacuum}},}\ }\href {\doibase
  10.1103/PhysRevD.104.054031} {\bibfield  {journal} {\bibinfo  {journal}
  {Phys. Rev. D}\ }\textbf {\bibinfo {volume} {104}},\ \bibinfo {pages}
  {054031} (\bibinfo {year} {2021})},\ \Eprint
  {http://arxiv.org/abs/2102.08191} {arXiv:2102.08191 [hep-ph]} \BibitemShut
  {NoStop}%
\bibitem [{\citenamefont {Leinweber}(1999)}]{Leinweber:1999cw}%
  \BibitemOpen
  \bibfield  {author} {\bibinfo {author} {\bibfnamefont {Derek~B.}\
  \bibnamefont {Leinweber}},\ }\bibfield  {title} {\enquote {\bibinfo {title}
  {{Visualizations of the QCD vacuum}},}\ }in\ \href@noop {} {\emph {\bibinfo
  {booktitle} {{Workshop on Light-Cone QCD and Nonperturbative Hadron
  Physics}}}}\ (\bibinfo {year} {1999})\ pp.\ \bibinfo {pages} {138--143},\
  \Eprint {http://arxiv.org/abs/hep-lat/0004025} {arXiv:hep-lat/0004025}
  \BibitemShut {NoStop}%
\bibitem [{\citenamefont {Biddle}\ \emph {et~al.}(2023)\citenamefont {Biddle},
  \citenamefont {Kamleh},\ and\ \citenamefont {Leinweber}}]{Biddle:2023lod}%
  \BibitemOpen
  \bibfield  {author} {\bibinfo {author} {\bibfnamefont {James~C.}\
  \bibnamefont {Biddle}}, \bibinfo {author} {\bibfnamefont {Waseem}\
  \bibnamefont {Kamleh}}, \ and\ \bibinfo {author} {\bibfnamefont {Derek~B.}\
  \bibnamefont {Leinweber}},\ }\bibfield  {title} {\enquote {\bibinfo {title}
  {{Center vortex structure in the presence of dynamical fermions}},}\ }\href
  {\doibase 10.1103/PhysRevD.107.094507} {\bibfield  {journal} {\bibinfo
  {journal} {Phys. Rev. D}\ }\textbf {\bibinfo {volume} {107}},\ \bibinfo
  {pages} {094507} (\bibinfo {year} {2023})},\ \Eprint
  {http://arxiv.org/abs/2302.05897} {arXiv:2302.05897 [hep-lat]} \BibitemShut
  {NoStop}%
\bibitem [{\citenamefont {Hafidi}\ \emph {et~al.}(2017)\citenamefont {Hafidi},
  \citenamefont {Joosten}, \citenamefont {Meziani},\ and\ \citenamefont
  {Qiu}}]{Hafidi:2017bsg}%
  \BibitemOpen
  \bibfield  {author} {\bibinfo {author} {\bibfnamefont {K.}~\bibnamefont
  {Hafidi}}, \bibinfo {author} {\bibfnamefont {S.}~\bibnamefont {Joosten}},
  \bibinfo {author} {\bibfnamefont {Z.~E.}\ \bibnamefont {Meziani}}, \ and\
  \bibinfo {author} {\bibfnamefont {J.~W.}\ \bibnamefont {Qiu}},\ }\bibfield
  {title} {\enquote {\bibinfo {title} {{Production of Charmonium at Threshold
  in Hall A and C at Jefferson Lab}},}\ }\href {\doibase
  10.1007/s00601-017-1305-3} {\bibfield  {journal} {\bibinfo  {journal} {Few
  Body Syst.}\ }\textbf {\bibinfo {volume} {58}},\ \bibinfo {pages} {141}
  (\bibinfo {year} {2017})}\BibitemShut {NoStop}%
\bibitem [{\citenamefont {Ali}\ \emph {et~al.}(2019)\citenamefont {Ali} \emph
  {et~al.}}]{GlueX:2019mkq}%
  \BibitemOpen
  \bibfield  {author} {\bibinfo {author} {\bibfnamefont {A.}~\bibnamefont
  {Ali}} \emph {et~al.} (\bibinfo {collaboration} {GlueX}),\ }\bibfield
  {title} {\enquote {\bibinfo {title} {{First Measurement of Near-Threshold
  J/\ensuremath{\psi} Exclusive Photoproduction off the Proton}},}\ }\href
  {\doibase 10.1103/PhysRevLett.123.072001} {\bibfield  {journal} {\bibinfo
  {journal} {Phys. Rev. Lett.}\ }\textbf {\bibinfo {volume} {123}},\ \bibinfo
  {pages} {072001} (\bibinfo {year} {2019})},\ \Eprint
  {http://arxiv.org/abs/1905.10811} {arXiv:1905.10811 [nucl-ex]} \BibitemShut
  {NoStop}%
\bibitem [{\citenamefont {Meziani}\ and\ \citenamefont
  {Joosten}(2020)}]{Meziani:2020oks}%
  \BibitemOpen
  \bibfield  {author} {\bibinfo {author} {\bibfnamefont {Zein-Eddine}\
  \bibnamefont {Meziani}}\ and\ \bibinfo {author} {\bibfnamefont {Sylvester}\
  \bibnamefont {Joosten}},\ }\bibfield  {title} {\enquote {\bibinfo {title}
  {{Origin of the Proton Mass? Heavy Quarkonium Production at Threshold from
  Jefferson Lab to an Electron Ion Collider}},}\ }in\ \href {\doibase
  10.1142/9789811214950_0048} {\emph {\bibinfo {booktitle} {{Probing Nucleons
  and Nuclei in High Energy Collisions}: {Dedicated to the Physics of the
  Electron Ion Collider}}}}\ (\bibinfo {year} {2020})\ pp.\ \bibinfo {pages}
  {234--237}\BibitemShut {NoStop}%
\bibitem [{\citenamefont {Anderle}\ \emph {et~al.}(2021)\citenamefont {Anderle}
  \emph {et~al.}}]{Anderle:2021wcy}%
  \BibitemOpen
  \bibfield  {author} {\bibinfo {author} {\bibfnamefont {Daniele~P.}\
  \bibnamefont {Anderle}} \emph {et~al.},\ }\bibfield  {title} {\enquote
  {\bibinfo {title} {{Electron-ion collider in China}},}\ }\href {\doibase
  10.1007/s11467-021-1062-0} {\bibfield  {journal} {\bibinfo  {journal} {Front.
  Phys. (Beijing)}\ }\textbf {\bibinfo {volume} {16}},\ \bibinfo {pages}
  {64701} (\bibinfo {year} {2021})},\ \Eprint {http://arxiv.org/abs/2102.09222}
  {arXiv:2102.09222 [nucl-ex]} \BibitemShut {NoStop}%
\bibitem [{\citenamefont {Hatta}\ and\ \citenamefont
  {Yang}(2018)}]{Hatta:2018ina}%
  \BibitemOpen
  \bibfield  {author} {\bibinfo {author} {\bibfnamefont {Yoshitaka}\
  \bibnamefont {Hatta}}\ and\ \bibinfo {author} {\bibfnamefont {Di-Lun}\
  \bibnamefont {Yang}},\ }\bibfield  {title} {\enquote {\bibinfo {title}
  {{Holographic $J/\psi$ production near threshold and the proton mass
  problem}},}\ }\href {\doibase 10.1103/PhysRevD.98.074003} {\bibfield
  {journal} {\bibinfo  {journal} {Phys. Rev. D}\ }\textbf {\bibinfo {volume}
  {98}},\ \bibinfo {pages} {074003} (\bibinfo {year} {2018})},\ \Eprint
  {http://arxiv.org/abs/1808.02163} {arXiv:1808.02163 [hep-ph]} \BibitemShut
  {NoStop}%
\bibitem [{\citenamefont {Mamo}\ and\ \citenamefont
  {Zahed}(2020)}]{Mamo:2019mka}%
  \BibitemOpen
  \bibfield  {author} {\bibinfo {author} {\bibfnamefont {Kiminad~A.}\
  \bibnamefont {Mamo}}\ and\ \bibinfo {author} {\bibfnamefont {Ismail}\
  \bibnamefont {Zahed}},\ }\bibfield  {title} {\enquote {\bibinfo {title}
  {{Diffractive photoproduction of $J/\psi$ and $\Upsilon$ using holographic
  QCD: gravitational form factors and GPD of gluons in the proton}},}\ }\href
  {\doibase 10.1103/PhysRevD.101.086003} {\bibfield  {journal} {\bibinfo
  {journal} {Phys. Rev. D}\ }\textbf {\bibinfo {volume} {101}},\ \bibinfo
  {pages} {086003} (\bibinfo {year} {2020})},\ \Eprint
  {http://arxiv.org/abs/1910.04707} {arXiv:1910.04707 [hep-ph]} \BibitemShut
  {NoStop}%
\bibitem [{\citenamefont {Kharzeev}(2021)}]{Kharzeev:2021qkd}%
  \BibitemOpen
  \bibfield  {author} {\bibinfo {author} {\bibfnamefont {Dmitri~E.}\
  \bibnamefont {Kharzeev}},\ }\bibfield  {title} {\enquote {\bibinfo {title}
  {{Mass radius of the proton}},}\ }\href {\doibase
  10.1103/PhysRevD.104.054015} {\bibfield  {journal} {\bibinfo  {journal}
  {Phys. Rev. D}\ }\textbf {\bibinfo {volume} {104}},\ \bibinfo {pages}
  {054015} (\bibinfo {year} {2021})},\ \Eprint
  {http://arxiv.org/abs/2102.00110} {arXiv:2102.00110 [hep-ph]} \BibitemShut
  {NoStop}%
\bibitem [{\citenamefont {Ji}(2021)}]{Ji:2021mtz}%
  \BibitemOpen
  \bibfield  {author} {\bibinfo {author} {\bibfnamefont {Xiangdong}\
  \bibnamefont {Ji}},\ }\bibfield  {title} {\enquote {\bibinfo {title} {{Proton
  mass decomposition: naturalness and interpretations}},}\ }\href {\doibase
  10.1007/s11467-021-1065-x} {\bibfield  {journal} {\bibinfo  {journal} {Front.
  Phys. (Beijing)}\ }\textbf {\bibinfo {volume} {16}},\ \bibinfo {pages}
  {64601} (\bibinfo {year} {2021})},\ \Eprint {http://arxiv.org/abs/2102.07830}
  {arXiv:2102.07830 [hep-ph]} \BibitemShut {NoStop}%
\bibitem [{\citenamefont {Hatta}\ and\ \citenamefont
  {Strikman}(2021)}]{Hatta:2021can}%
  \BibitemOpen
  \bibfield  {author} {\bibinfo {author} {\bibfnamefont {Yoshitaka}\
  \bibnamefont {Hatta}}\ and\ \bibinfo {author} {\bibfnamefont {Mark}\
  \bibnamefont {Strikman}},\ }\bibfield  {title} {\enquote {\bibinfo {title}
  {{$\phi$-meson lepto-production near threshold and the strangeness
  $D$-term}},}\ }\href {\doibase 10.1016/j.physletb.2021.136295} {\bibfield
  {journal} {\bibinfo  {journal} {Phys. Lett. B}\ }\textbf {\bibinfo {volume}
  {817}},\ \bibinfo {pages} {136295} (\bibinfo {year} {2021})},\ \Eprint
  {http://arxiv.org/abs/2102.12631} {arXiv:2102.12631 [hep-ph]} \BibitemShut
  {NoStop}%
\bibitem [{\citenamefont {Guo}\ \emph {et~al.}(2021{\natexlab{a}})\citenamefont
  {Guo}, \citenamefont {Ji},\ and\ \citenamefont {Liu}}]{Guo:2021ibg}%
  \BibitemOpen
  \bibfield  {author} {\bibinfo {author} {\bibfnamefont {Yuxun}\ \bibnamefont
  {Guo}}, \bibinfo {author} {\bibfnamefont {Xiangdong}\ \bibnamefont {Ji}}, \
  and\ \bibinfo {author} {\bibfnamefont {Yizhuang}\ \bibnamefont {Liu}},\
  }\bibfield  {title} {\enquote {\bibinfo {title} {{QCD Analysis of
  Near-Threshold Photon-Proton Production of Heavy Quarkonium}},}\ }\href
  {\doibase 10.1103/PhysRevD.103.096010} {\bibfield  {journal} {\bibinfo
  {journal} {Phys. Rev. D}\ }\textbf {\bibinfo {volume} {103}},\ \bibinfo
  {pages} {096010} (\bibinfo {year} {2021}{\natexlab{a}})},\ \Eprint
  {http://arxiv.org/abs/2103.11506} {arXiv:2103.11506 [hep-ph]} \BibitemShut
  {NoStop}%
\bibitem [{\citenamefont {Sun}\ \emph {et~al.}(2021)\citenamefont {Sun},
  \citenamefont {Tong},\ and\ \citenamefont {Yuan}}]{Sun:2021gmi}%
  \BibitemOpen
  \bibfield  {author} {\bibinfo {author} {\bibfnamefont {Peng}\ \bibnamefont
  {Sun}}, \bibinfo {author} {\bibfnamefont {Xuan-Bo}\ \bibnamefont {Tong}}, \
  and\ \bibinfo {author} {\bibfnamefont {Feng}\ \bibnamefont {Yuan}},\
  }\bibfield  {title} {\enquote {\bibinfo {title} {{Perturbative QCD analysis
  of near threshold heavy quarkonium photoproduction at large momentum
  transfer}},}\ }\href {\doibase 10.1016/j.physletb.2021.136655} {\bibfield
  {journal} {\bibinfo  {journal} {Phys. Lett. B}\ }\textbf {\bibinfo {volume}
  {822}},\ \bibinfo {pages} {136655} (\bibinfo {year} {2021})},\ \Eprint
  {http://arxiv.org/abs/2103.12047} {arXiv:2103.12047 [hep-ph]} \BibitemShut
  {NoStop}%
\bibitem [{\citenamefont {Wang}\ \emph {et~al.}(2022)\citenamefont {Wang},
  \citenamefont {Zeng},\ and\ \citenamefont {Wang}}]{Wang:2022vhr}%
  \BibitemOpen
  \bibfield  {author} {\bibinfo {author} {\bibfnamefont {Xiao-Yun}\
  \bibnamefont {Wang}}, \bibinfo {author} {\bibfnamefont {Fancong}\
  \bibnamefont {Zeng}}, \ and\ \bibinfo {author} {\bibfnamefont {Quanjin}\
  \bibnamefont {Wang}},\ }\bibfield  {title} {\enquote {\bibinfo {title}
  {{Systematic analysis of the proton mass radius based on photoproduction of
  vector charmoniums}},}\ }\href {\doibase 10.1103/PhysRevD.105.096033}
  {\bibfield  {journal} {\bibinfo  {journal} {Phys. Rev. D}\ }\textbf {\bibinfo
  {volume} {105}},\ \bibinfo {pages} {096033} (\bibinfo {year} {2022})},\
  \Eprint {http://arxiv.org/abs/2204.07294} {arXiv:2204.07294 [hep-ph]}
  \BibitemShut {NoStop}%
\bibitem [{\citenamefont {Bartels}(1980)}]{Bartels:1980pe}%
  \BibitemOpen
  \bibfield  {author} {\bibinfo {author} {\bibfnamefont {Jochen}\ \bibnamefont
  {Bartels}},\ }\bibfield  {title} {\enquote {\bibinfo {title} {{High-Energy
  Behavior in a Nonabelian Gauge Theory (II)}: {First Corrections to $T_{n\to
  m}$ Beyond the Leading $\ln s$ Approximation}},}\ }\href {\doibase
  10.1016/0550-3213(80)90019-X} {\bibfield  {journal} {\bibinfo  {journal}
  {Nucl. Phys. B}\ }\textbf {\bibinfo {volume} {175}},\ \bibinfo {pages}
  {365--401} (\bibinfo {year} {1980})}\BibitemShut {NoStop}%
\bibitem [{\citenamefont {Kwiecinski}\ and\ \citenamefont
  {Praszalowicz}(1980)}]{Kwiecinski:1980wb}%
  \BibitemOpen
  \bibfield  {author} {\bibinfo {author} {\bibfnamefont {J.}~\bibnamefont
  {Kwiecinski}}\ and\ \bibinfo {author} {\bibfnamefont {M.}~\bibnamefont
  {Praszalowicz}},\ }\bibfield  {title} {\enquote {\bibinfo {title} {{Three
  Gluon Integral Equation and Odd c Singlet Regge Singularities in QCD}},}\
  }\href {\doibase 10.1016/0370-2693(80)90909-0} {\bibfield  {journal}
  {\bibinfo  {journal} {Phys. Lett. B}\ }\textbf {\bibinfo {volume} {94}},\
  \bibinfo {pages} {413--416} (\bibinfo {year} {1980})}\BibitemShut {NoStop}%
\bibitem [{\citenamefont {Braun}(1998)}]{Braun:1998fs}%
  \BibitemOpen
  \bibfield  {author} {\bibinfo {author} {\bibfnamefont {M.~A.}\ \bibnamefont
  {Braun}},\ }\bibfield  {title} {\enquote {\bibinfo {title} {{Odderon and
  QCD}},}\ }\href@noop {} {\  (\bibinfo {year} {1998})},\ \Eprint
  {http://arxiv.org/abs/hep-ph/9805394} {arXiv:hep-ph/9805394} \BibitemShut
  {NoStop}%
\bibitem [{\citenamefont {Abazov}\ \emph {et~al.}(2021)\citenamefont {Abazov}
  \emph {et~al.}}]{TOTEM:2020zzr}%
  \BibitemOpen
  \bibfield  {author} {\bibinfo {author} {\bibfnamefont {V.~M.}\ \bibnamefont
  {Abazov}} \emph {et~al.} (\bibinfo {collaboration} {TOTEM, D0}),\ }\bibfield
  {title} {\enquote {\bibinfo {title} {{Odderon Exchange from Elastic
  Scattering Differences between $pp$ and $p \bar{p}$ Data at 1.96~TeV and from
  pp Forward Scattering Measurements}},}\ }\href {\doibase
  10.1103/PhysRevLett.127.062003} {\bibfield  {journal} {\bibinfo  {journal}
  {Phys. Rev. Lett.}\ }\textbf {\bibinfo {volume} {127}},\ \bibinfo {pages}
  {062003} (\bibinfo {year} {2021})},\ \Eprint
  {http://arxiv.org/abs/2012.03981} {arXiv:2012.03981 [hep-ex]} \BibitemShut
  {NoStop}%
\bibitem [{\citenamefont {Hechenberger}\ \emph {et~al.}(2024)\citenamefont
  {Hechenberger}, \citenamefont {Mamo},\ and\ \citenamefont
  {Zahed}}]{Hechenberger:2024abg}%
  \BibitemOpen
  \bibfield  {author} {\bibinfo {author} {\bibfnamefont {Florian}\ \bibnamefont
  {Hechenberger}}, \bibinfo {author} {\bibfnamefont {Kiminad~A.}\ \bibnamefont
  {Mamo}}, \ and\ \bibinfo {author} {\bibfnamefont {Ismail}\ \bibnamefont
  {Zahed}},\ }\bibfield  {title} {\enquote {\bibinfo {title} {{Threshold
  production of $\eta_{c,b}$ using holographic QCD}},}\ }\href@noop {} {\
  (\bibinfo {year} {2024})},\ \Eprint {http://arxiv.org/abs/2401.12162}
  {arXiv:2401.12162 [hep-ph]} \BibitemShut {NoStop}%
\bibitem [{\citenamefont {Sun}\ \emph {et~al.}(2022)\citenamefont {Sun},
  \citenamefont {Tong},\ and\ \citenamefont {Yuan}}]{Sun:2021pyw}%
  \BibitemOpen
  \bibfield  {author} {\bibinfo {author} {\bibfnamefont {Peng}\ \bibnamefont
  {Sun}}, \bibinfo {author} {\bibfnamefont {Xuan-Bo}\ \bibnamefont {Tong}}, \
  and\ \bibinfo {author} {\bibfnamefont {Feng}\ \bibnamefont {Yuan}},\
  }\bibfield  {title} {\enquote {\bibinfo {title} {{Near threshold heavy
  quarkonium photoproduction at large momentum transfer}},}\ }\href {\doibase
  10.1103/PhysRevD.105.054032} {\bibfield  {journal} {\bibinfo  {journal}
  {Phys. Rev. D}\ }\textbf {\bibinfo {volume} {105}},\ \bibinfo {pages}
  {054032} (\bibinfo {year} {2022})},\ \Eprint
  {http://arxiv.org/abs/2111.07034} {arXiv:2111.07034 [hep-ph]} \BibitemShut
  {NoStop}%
\bibitem [{\citenamefont {Ma}(2003)}]{Ma:2003py}%
  \BibitemOpen
  \bibfield  {author} {\bibinfo {author} {\bibfnamefont {J.~P.}\ \bibnamefont
  {Ma}},\ }\bibfield  {title} {\enquote {\bibinfo {title} {{Diffractive
  photoproduction of eta(c)}},}\ }\href {\doibase
  10.1016/j.nuclphysa.2003.08.016} {\bibfield  {journal} {\bibinfo  {journal}
  {Nucl. Phys. A}\ }\textbf {\bibinfo {volume} {727}},\ \bibinfo {pages}
  {333--352} (\bibinfo {year} {2003})},\ \Eprint
  {http://arxiv.org/abs/hep-ph/0301155} {arXiv:hep-ph/0301155} \BibitemShut
  {NoStop}%
\bibitem [{\citenamefont {Ji}\ \emph {et~al.}(2004)\citenamefont {Ji},
  \citenamefont {Ma},\ and\ \citenamefont {Yuan}}]{Ji:2003yj}%
  \BibitemOpen
  \bibfield  {author} {\bibinfo {author} {\bibfnamefont {Xiang-dong}\
  \bibnamefont {Ji}}, \bibinfo {author} {\bibfnamefont {Jian-Ping}\
  \bibnamefont {Ma}}, \ and\ \bibinfo {author} {\bibfnamefont {Feng}\
  \bibnamefont {Yuan}},\ }\bibfield  {title} {\enquote {\bibinfo {title}
  {{Classification and asymptotic scaling of hadrons' light cone wave function
  amplitudes}},}\ }\href {\doibase 10.1140/epjc/s2003-01563-y} {\bibfield
  {journal} {\bibinfo  {journal} {Eur. Phys. J. C}\ }\textbf {\bibinfo {volume}
  {33}},\ \bibinfo {pages} {75--90} (\bibinfo {year} {2004})},\ \Eprint
  {http://arxiv.org/abs/hep-ph/0304107} {arXiv:hep-ph/0304107} \BibitemShut
  {NoStop}%
\bibitem [{\citenamefont {Ji}(1998)}]{Ji:1998pc}%
  \BibitemOpen
  \bibfield  {author} {\bibinfo {author} {\bibfnamefont {Xiang-Dong}\
  \bibnamefont {Ji}},\ }\bibfield  {title} {\enquote {\bibinfo {title} {{Off
  forward parton distributions}},}\ }\href {\doibase
  10.1088/0954-3899/24/7/002} {\bibfield  {journal} {\bibinfo  {journal} {J.
  Phys. G}\ }\textbf {\bibinfo {volume} {24}},\ \bibinfo {pages} {1181--1205}
  (\bibinfo {year} {1998})},\ \Eprint {http://arxiv.org/abs/hep-ph/9807358}
  {arXiv:hep-ph/9807358} \BibitemShut {NoStop}%
\bibitem [{\citenamefont {Guo}\ \emph {et~al.}(2021{\natexlab{b}})\citenamefont
  {Guo}, \citenamefont {Ji},\ and\ \citenamefont {Shiells}}]{Guo:2021aik}%
  \BibitemOpen
  \bibfield  {author} {\bibinfo {author} {\bibfnamefont {Yuxun}\ \bibnamefont
  {Guo}}, \bibinfo {author} {\bibfnamefont {Xiangdong}\ \bibnamefont {Ji}}, \
  and\ \bibinfo {author} {\bibfnamefont {Kyle}\ \bibnamefont {Shiells}},\
  }\bibfield  {title} {\enquote {\bibinfo {title} {{Novel twist-three
  transverse-spin sum rule for the proton and related generalized parton
  distributions}},}\ }\href {\doibase 10.1016/j.nuclphysb.2021.115440}
  {\bibfield  {journal} {\bibinfo  {journal} {Nucl. Phys. B}\ }\textbf
  {\bibinfo {volume} {969}},\ \bibinfo {pages} {115440} (\bibinfo {year}
  {2021}{\natexlab{b}})},\ \Eprint {http://arxiv.org/abs/2101.05243}
  {arXiv:2101.05243 [hep-ph]} \BibitemShut {NoStop}%
\bibitem [{\citenamefont {Diehl}(2001)}]{Diehl:2001pm}%
  \BibitemOpen
  \bibfield  {author} {\bibinfo {author} {\bibfnamefont {M.}~\bibnamefont
  {Diehl}},\ }\bibfield  {title} {\enquote {\bibinfo {title} {{Generalized
  parton distributions with helicity flip}},}\ }\href {\doibase
  10.1007/s100520100635} {\bibfield  {journal} {\bibinfo  {journal} {Eur. Phys.
  J. C}\ }\textbf {\bibinfo {volume} {19}},\ \bibinfo {pages} {485--492}
  (\bibinfo {year} {2001})},\ \Eprint {http://arxiv.org/abs/hep-ph/0101335}
  {arXiv:hep-ph/0101335} \BibitemShut {NoStop}%
\bibitem [{\citenamefont {Liu}\ \emph {et~al.}(2024)\citenamefont {Liu},
  \citenamefont {Shuryak},\ and\ \citenamefont {Zahed}}]{Liu:2024}%
  \BibitemOpen
  \bibfield  {author} {\bibinfo {author} {\bibfnamefont {Wei-Yang}\
  \bibnamefont {Liu}}, \bibinfo {author} {\bibfnamefont {Edward}\ \bibnamefont
  {Shuryak}}, \ and\ \bibinfo {author} {\bibfnamefont {Ismail}\ \bibnamefont
  {Zahed}},\ }\bibfield  {title} {\enquote {\bibinfo {title} {{Glue in hadrons
  at medium resolution and the QCD instanton vacuum}},}\ }\href@noop {} {\
  (\bibinfo {year} {2024})},\ \Eprint {http://arxiv.org/abs/2404.03047}
  {arXiv:2404.03047 [hep-ph]} \BibitemShut {NoStop}%
\bibitem [{\citenamefont {Ji}\ and\ \citenamefont {Lebed}(2001)}]{Ji:2000id}%
  \BibitemOpen
  \bibfield  {author} {\bibinfo {author} {\bibfnamefont {Xiang-Dong}\
  \bibnamefont {Ji}}\ and\ \bibinfo {author} {\bibfnamefont {Richard~F.}\
  \bibnamefont {Lebed}},\ }\bibfield  {title} {\enquote {\bibinfo {title}
  {{Counting form-factors of twist-two operators}},}\ }\href {\doibase
  10.1103/PhysRevD.63.076005} {\bibfield  {journal} {\bibinfo  {journal} {Phys.
  Rev. D}\ }\textbf {\bibinfo {volume} {63}},\ \bibinfo {pages} {076005}
  (\bibinfo {year} {2001})},\ \Eprint {http://arxiv.org/abs/hep-ph/0012160}
  {arXiv:hep-ph/0012160} \BibitemShut {NoStop}%
\bibitem [{\citenamefont {Polyakov}\ and\ \citenamefont
  {Sun}(2019)}]{Polyakov:2019lbq}%
  \BibitemOpen
  \bibfield  {author} {\bibinfo {author} {\bibfnamefont {Maxim~V.}\
  \bibnamefont {Polyakov}}\ and\ \bibinfo {author} {\bibfnamefont {Bao-Dong}\
  \bibnamefont {Sun}},\ }\bibfield  {title} {\enquote {\bibinfo {title}
  {{Gravitational form factors of a spin one particle}},}\ }\href {\doibase
  10.1103/PhysRevD.100.036003} {\bibfield  {journal} {\bibinfo  {journal}
  {Phys. Rev. D}\ }\textbf {\bibinfo {volume} {100}},\ \bibinfo {pages}
  {036003} (\bibinfo {year} {2019})},\ \Eprint
  {http://arxiv.org/abs/1903.02738} {arXiv:1903.02738 [hep-ph]} \BibitemShut
  {NoStop}%
\bibitem [{\citenamefont {Hou}\ \emph {et~al.}(2021)\citenamefont {Hou} \emph
  {et~al.}}]{Hou:2019efy}%
  \BibitemOpen
  \bibfield  {author} {\bibinfo {author} {\bibfnamefont {Tie-Jiun}\
  \bibnamefont {Hou}} \emph {et~al.},\ }\bibfield  {title} {\enquote {\bibinfo
  {title} {{New CTEQ global analysis of quantum chromodynamics with
  high-precision data from the LHC}},}\ }\href {\doibase
  10.1103/PhysRevD.103.014013} {\bibfield  {journal} {\bibinfo  {journal}
  {Phys. Rev. D}\ }\textbf {\bibinfo {volume} {103}},\ \bibinfo {pages}
  {014013} (\bibinfo {year} {2021})},\ \Eprint
  {http://arxiv.org/abs/1912.10053} {arXiv:1912.10053 [hep-ph]} \BibitemShut
  {NoStop}%
\bibitem [{\citenamefont {Guo}\ \emph {et~al.}(2023)\citenamefont {Guo},
  \citenamefont {Ji}, \citenamefont {Liu},\ and\ \citenamefont
  {Yang}}]{Guo:2023pqw}%
  \BibitemOpen
  \bibfield  {author} {\bibinfo {author} {\bibfnamefont {Yuxun}\ \bibnamefont
  {Guo}}, \bibinfo {author} {\bibfnamefont {Xiangdong}\ \bibnamefont {Ji}},
  \bibinfo {author} {\bibfnamefont {Yizhuang}\ \bibnamefont {Liu}}, \ and\
  \bibinfo {author} {\bibfnamefont {Jinghong}\ \bibnamefont {Yang}},\
  }\bibfield  {title} {\enquote {\bibinfo {title} {{Updated analysis of
  near-threshold heavy quarkonium production for probe of
  proton\textquoteright{}s gluonic gravitational form factors}},}\ }\href
  {\doibase 10.1103/PhysRevD.108.034003} {\bibfield  {journal} {\bibinfo
  {journal} {Phys. Rev. D}\ }\textbf {\bibinfo {volume} {108}},\ \bibinfo
  {pages} {034003} (\bibinfo {year} {2023})},\ \Eprint
  {http://arxiv.org/abs/2305.06992} {arXiv:2305.06992 [hep-ph]} \BibitemShut
  {NoStop}%
\bibitem [{\citenamefont {Hackett}\ \emph {et~al.}(2023)\citenamefont
  {Hackett}, \citenamefont {Pefkou},\ and\ \citenamefont
  {Shanahan}}]{Hackett:2023rif}%
  \BibitemOpen
  \bibfield  {author} {\bibinfo {author} {\bibfnamefont {Daniel~C.}\
  \bibnamefont {Hackett}}, \bibinfo {author} {\bibfnamefont {Dimitra~A.}\
  \bibnamefont {Pefkou}}, \ and\ \bibinfo {author} {\bibfnamefont {Phiala~E.}\
  \bibnamefont {Shanahan}},\ }\bibfield  {title} {\enquote {\bibinfo {title}
  {{Gravitational form factors of the proton from lattice QCD}},}\ }\href@noop
  {} {\  (\bibinfo {year} {2023})},\ \Eprint {http://arxiv.org/abs/2310.08484}
  {arXiv:2310.08484 [hep-lat]} \BibitemShut {NoStop}%
\bibitem [{\citenamefont {Pefkou}\ \emph {et~al.}(2022)\citenamefont {Pefkou},
  \citenamefont {Hackett},\ and\ \citenamefont {Shanahan}}]{Pefkou:2021fni}%
  \BibitemOpen
  \bibfield  {author} {\bibinfo {author} {\bibfnamefont {Dimitra~A.}\
  \bibnamefont {Pefkou}}, \bibinfo {author} {\bibfnamefont {Daniel~C.}\
  \bibnamefont {Hackett}}, \ and\ \bibinfo {author} {\bibfnamefont {Phiala~E.}\
  \bibnamefont {Shanahan}},\ }\bibfield  {title} {\enquote {\bibinfo {title}
  {{Gluon gravitational structure of hadrons of different spin}},}\ }\href
  {\doibase 10.1103/PhysRevD.105.054509} {\bibfield  {journal} {\bibinfo
  {journal} {Phys. Rev. D}\ }\textbf {\bibinfo {volume} {105}},\ \bibinfo
  {pages} {054509} (\bibinfo {year} {2022})},\ \Eprint
  {http://arxiv.org/abs/2107.10368} {arXiv:2107.10368 [hep-lat]} \BibitemShut
  {NoStop}%
\bibitem [{\citenamefont {Ye}\ \emph {et~al.}(2017)\citenamefont {Ye},
  \citenamefont {Sato}, \citenamefont {Allada}, \citenamefont {Liu},
  \citenamefont {Chen}, \citenamefont {Gao}, \citenamefont {Kang},
  \citenamefont {Prokudin}, \citenamefont {Sun},\ and\ \citenamefont
  {Yuan}}]{YE201791}%
  \BibitemOpen
  \bibfield  {author} {\bibinfo {author} {\bibfnamefont {Zhihong}\ \bibnamefont
  {Ye}}, \bibinfo {author} {\bibfnamefont {Nobuo}\ \bibnamefont {Sato}},
  \bibinfo {author} {\bibfnamefont {Kalyan}\ \bibnamefont {Allada}}, \bibinfo
  {author} {\bibfnamefont {Tianbo}\ \bibnamefont {Liu}}, \bibinfo {author}
  {\bibfnamefont {Jian-Ping}\ \bibnamefont {Chen}}, \bibinfo {author}
  {\bibfnamefont {Haiyan}\ \bibnamefont {Gao}}, \bibinfo {author}
  {\bibfnamefont {Zhong-Bo}\ \bibnamefont {Kang}}, \bibinfo {author}
  {\bibfnamefont {Alexei}\ \bibnamefont {Prokudin}}, \bibinfo {author}
  {\bibfnamefont {Peng}\ \bibnamefont {Sun}}, \ and\ \bibinfo {author}
  {\bibfnamefont {Feng}\ \bibnamefont {Yuan}},\ }\bibfield  {title} {\enquote
  {\bibinfo {title} {Unveiling the nucleon tensor charge at jefferson lab: A
  study of the solid case},}\ }\href {\doibase
  https://doi.org/10.1016/j.physletb.2017.01.046} {\bibfield  {journal}
  {\bibinfo  {journal} {Physics Letters B}\ }\textbf {\bibinfo {volume}
  {767}},\ \bibinfo {pages} {91--98} (\bibinfo {year} {2017})}\BibitemShut
  {NoStop}%
\bibitem [{\citenamefont {Kang}\ \emph {et~al.}(2016)\citenamefont {Kang},
  \citenamefont {Prokudin}, \citenamefont {Sun},\ and\ \citenamefont
  {Yuan}}]{Kang:2015msa}%
  \BibitemOpen
  \bibfield  {author} {\bibinfo {author} {\bibfnamefont {Zhong-Bo}\
  \bibnamefont {Kang}}, \bibinfo {author} {\bibfnamefont {Alexei}\ \bibnamefont
  {Prokudin}}, \bibinfo {author} {\bibfnamefont {Peng}\ \bibnamefont {Sun}}, \
  and\ \bibinfo {author} {\bibfnamefont {Feng}\ \bibnamefont {Yuan}},\
  }\bibfield  {title} {\enquote {\bibinfo {title} {{Extraction of Quark
  Transversity Distribution and Collins Fragmentation Functions with QCD
  Evolution}},}\ }\href {\doibase 10.1103/PhysRevD.93.014009} {\bibfield
  {journal} {\bibinfo  {journal} {Phys. Rev. D}\ }\textbf {\bibinfo {volume}
  {93}},\ \bibinfo {pages} {014009} (\bibinfo {year} {2016})},\ \Eprint
  {http://arxiv.org/abs/1505.05589} {arXiv:1505.05589 [hep-ph]} \BibitemShut
  {NoStop}%
\bibitem [{\citenamefont {Artru}\ and\ \citenamefont
  {Mekhfi}(1990)}]{artru1990transversely}%
  \BibitemOpen
  \bibfield  {author} {\bibinfo {author} {\bibfnamefont {Xavier}\ \bibnamefont
  {Artru}}\ and\ \bibinfo {author} {\bibfnamefont {Mustapha}\ \bibnamefont
  {Mekhfi}},\ }\bibfield  {title} {\enquote {\bibinfo {title} {Transversely
  polarized parton densities, their evolution and their measurement},}\
  }\href@noop {} {\bibfield  {journal} {\bibinfo  {journal} {Zeitschrift
  f{\"u}r Physik C Particles and Fields}\ }\textbf {\bibinfo {volume} {45}},\
  \bibinfo {pages} {669--676} (\bibinfo {year} {1990})}\BibitemShut {NoStop}%
\bibitem [{\citenamefont {Gamberg}\ and\ \citenamefont
  {Goldstein}(2001)}]{Gamberg:2001xc}%
  \BibitemOpen
  \bibfield  {author} {\bibinfo {author} {\bibfnamefont {Leonard~P.}\
  \bibnamefont {Gamberg}}\ and\ \bibinfo {author} {\bibfnamefont {Gary~R.}\
  \bibnamefont {Goldstein}},\ }\bibfield  {title} {\enquote {\bibinfo {title}
  {{Estimates of the nucleon tensor charge}},}\ }in\ \href@noop {} {\emph
  {\bibinfo {booktitle} {{20th International Symposium on Lepton and Photon
  Interactions at High Energies (LP 01)}}}}\ (\bibinfo {year} {2001})\ \Eprint
  {http://arxiv.org/abs/hep-ph/0106178} {arXiv:hep-ph/0106178} \BibitemShut
  {NoStop}%
\bibitem [{\citenamefont {Shanahan}\ and\ \citenamefont
  {Detmold}(2019)}]{Shanahan:2018pib}%
  \BibitemOpen
  \bibfield  {author} {\bibinfo {author} {\bibfnamefont {P.~E.}\ \bibnamefont
  {Shanahan}}\ and\ \bibinfo {author} {\bibfnamefont {W.}~\bibnamefont
  {Detmold}},\ }\bibfield  {title} {\enquote {\bibinfo {title} {{Gluon
  gravitational form factors of the nucleon and the pion from lattice QCD}},}\
  }\href {\doibase 10.1103/PhysRevD.99.014511} {\bibfield  {journal} {\bibinfo
  {journal} {Phys. Rev. D}\ }\textbf {\bibinfo {volume} {99}},\ \bibinfo
  {pages} {014511} (\bibinfo {year} {2019})},\ \Eprint
  {http://arxiv.org/abs/1810.04626} {arXiv:1810.04626 [hep-lat]} \BibitemShut
  {NoStop}%
\bibitem [{\citenamefont {Mamo}\ and\ \citenamefont
  {Zahed}(2021)}]{Mamo:2021krl}%
  \BibitemOpen
  \bibfield  {author} {\bibinfo {author} {\bibfnamefont {Kiminad~A.}\
  \bibnamefont {Mamo}}\ and\ \bibinfo {author} {\bibfnamefont {Ismail}\
  \bibnamefont {Zahed}},\ }\bibfield  {title} {\enquote {\bibinfo {title}
  {{Nucleon mass radii and distribution: Holographic QCD, Lattice QCD and GlueX
  data}},}\ }\href {\doibase 10.1103/PhysRevD.103.094010} {\bibfield  {journal}
  {\bibinfo  {journal} {Phys. Rev. D}\ }\textbf {\bibinfo {volume} {103}},\
  \bibinfo {pages} {094010} (\bibinfo {year} {2021})},\ \Eprint
  {http://arxiv.org/abs/2103.03186} {arXiv:2103.03186 [hep-ph]} \BibitemShut
  {NoStop}%
\bibitem [{\citenamefont {Mamo}\ and\ \citenamefont
  {Zahed}(2022)}]{Mamo:2022eui}%
  \BibitemOpen
  \bibfield  {author} {\bibinfo {author} {\bibfnamefont {Kiminad~A.}\
  \bibnamefont {Mamo}}\ and\ \bibinfo {author} {\bibfnamefont {Ismail}\
  \bibnamefont {Zahed}},\ }\bibfield  {title} {\enquote {\bibinfo {title}
  {{J/\ensuremath{\psi} near threshold in holographic QCD: A and D
  gravitational form factors}},}\ }\href {\doibase 10.1103/PhysRevD.106.086004}
  {\bibfield  {journal} {\bibinfo  {journal} {Phys. Rev. D}\ }\textbf {\bibinfo
  {volume} {106}},\ \bibinfo {pages} {086004} (\bibinfo {year} {2022})},\
  \Eprint {http://arxiv.org/abs/2204.08857} {arXiv:2204.08857 [hep-ph]}
  \BibitemShut {NoStop}%
\bibitem [{\citenamefont {Adhikari}\ \emph {et~al.}(2023)\citenamefont
  {Adhikari} \emph {et~al.}}]{GlueX:2023pev}%
  \BibitemOpen
  \bibfield  {author} {\bibinfo {author} {\bibfnamefont {S.}~\bibnamefont
  {Adhikari}} \emph {et~al.} (\bibinfo {collaboration} {GlueX}),\ }\bibfield
  {title} {\enquote {\bibinfo {title} {{Measurement of the J/$\psi $
  photoproduction cross section over the full near-threshold kinematic
  region}},}\ }\href {\doibase 10.1103/PhysRevC.108.025201} {\bibfield
  {journal} {\bibinfo  {journal} {Phys. Rev. C}\ }\textbf {\bibinfo {volume}
  {108}},\ \bibinfo {pages} {025201} (\bibinfo {year} {2023})},\ \Eprint
  {http://arxiv.org/abs/2304.03845} {arXiv:2304.03845 [nucl-ex]} \BibitemShut
  {NoStop}%
\bibitem [{\citenamefont {Duran}\ \emph {et~al.}(2023)\citenamefont {Duran}
  \emph {et~al.}}]{Duran:2022xag}%
  \BibitemOpen
  \bibfield  {author} {\bibinfo {author} {\bibfnamefont {B.}~\bibnamefont
  {Duran}} \emph {et~al.},\ }\bibfield  {title} {\enquote {\bibinfo {title}
  {{Determining the gluonic gravitational form factors of the proton}},}\
  }\href {\doibase 10.1038/s41586-023-05730-4} {\bibfield  {journal} {\bibinfo
  {journal} {Nature}\ }\textbf {\bibinfo {volume} {615}},\ \bibinfo {pages}
  {813--816} (\bibinfo {year} {2023})},\ \Eprint
  {http://arxiv.org/abs/2207.05212} {arXiv:2207.05212 [nucl-ex]} \BibitemShut
  {NoStop}%
\bibitem [{\citenamefont {Tanabashi}\ \emph {et~al.}(2018)\citenamefont
  {Tanabashi} \emph {et~al.}}]{ParticleDataGroup:2018ovx}%
  \BibitemOpen
  \bibfield  {author} {\bibinfo {author} {\bibfnamefont {M.}~\bibnamefont
  {Tanabashi}} \emph {et~al.} (\bibinfo {collaboration} {Particle Data
  Group}),\ }\bibfield  {title} {\enquote {\bibinfo {title} {{Review of
  Particle Physics}},}\ }\href {\doibase 10.1103/PhysRevD.98.030001} {\bibfield
   {journal} {\bibinfo  {journal} {Phys. Rev. D}\ }\textbf {\bibinfo {volume}
  {98}},\ \bibinfo {pages} {030001} (\bibinfo {year} {2018})}\BibitemShut
  {NoStop}%
\bibitem [{\citenamefont {Eidelman}\ \emph {et~al.}(2004)\citenamefont
  {Eidelman} \emph {et~al.}}]{ParticleDataGroup:2004fcd}%
  \BibitemOpen
  \bibfield  {author} {\bibinfo {author} {\bibfnamefont {S.}~\bibnamefont
  {Eidelman}} \emph {et~al.} (\bibinfo {collaboration} {Particle Data Group}),\
  }\bibfield  {title} {\enquote {\bibinfo {title} {{Review of particle physics.
  Particle Data Group}},}\ }\href {\doibase 10.1016/j.physletb.2004.06.001}
  {\bibfield  {journal} {\bibinfo  {journal} {Phys. Lett. B}\ }\textbf
  {\bibinfo {volume} {592}},\ \bibinfo {pages} {1} (\bibinfo {year}
  {2004})}\BibitemShut {NoStop}%
\bibitem [{\citenamefont {Olive}\ \emph {et~al.}(2014)\citenamefont {Olive}
  \emph {et~al.}}]{ParticleDataGroup:2014cgo}%
  \BibitemOpen
  \bibfield  {author} {\bibinfo {author} {\bibfnamefont {K.~A.}\ \bibnamefont
  {Olive}} \emph {et~al.} (\bibinfo {collaboration} {Particle Data Group}),\
  }\bibfield  {title} {\enquote {\bibinfo {title} {{Review of Particle
  Physics}},}\ }\href {\doibase 10.1088/1674-1137/38/9/090001} {\bibfield
  {journal} {\bibinfo  {journal} {Chin. Phys. C}\ }\textbf {\bibinfo {volume}
  {38}},\ \bibinfo {pages} {090001} (\bibinfo {year} {2014})}\BibitemShut
  {NoStop}%
\bibitem [{\citenamefont {Dumitru}\ and\ \citenamefont
  {Stebel}(2019)}]{Dumitru:2019qec}%
  \BibitemOpen
  \bibfield  {author} {\bibinfo {author} {\bibfnamefont {Adrian}\ \bibnamefont
  {Dumitru}}\ and\ \bibinfo {author} {\bibfnamefont {Tomasz}\ \bibnamefont
  {Stebel}},\ }\bibfield  {title} {\enquote {\bibinfo {title} {{Multiquark
  matrix elements in the proton and three gluon exchange for exclusive $\eta_c$
  production in photon-proton diffractive scattering}},}\ }\href {\doibase
  10.1103/PhysRevD.99.094038} {\bibfield  {journal} {\bibinfo  {journal} {Phys.
  Rev. D}\ }\textbf {\bibinfo {volume} {99}},\ \bibinfo {pages} {094038}
  (\bibinfo {year} {2019})},\ \Eprint {http://arxiv.org/abs/1903.07660}
  {arXiv:1903.07660 [hep-ph]} \BibitemShut {NoStop}%
\bibitem [{\citenamefont {Jia}\ \emph {et~al.}(2023)\citenamefont {Jia},
  \citenamefont {Mo}, \citenamefont {Pan},\ and\ \citenamefont
  {Zhang}}]{Jia:2022oyl}%
  \BibitemOpen
  \bibfield  {author} {\bibinfo {author} {\bibfnamefont {Yu}~\bibnamefont
  {Jia}}, \bibinfo {author} {\bibfnamefont {Zhewen}\ \bibnamefont {Mo}},
  \bibinfo {author} {\bibfnamefont {Jichen}\ \bibnamefont {Pan}}, \ and\
  \bibinfo {author} {\bibfnamefont {Jia-Yue}\ \bibnamefont {Zhang}},\
  }\bibfield  {title} {\enquote {\bibinfo {title} {{Photoproduction of C-even
  quarkonia at the EIC and EicC}},}\ }\href {\doibase
  10.1103/PhysRevD.108.016015} {\bibfield  {journal} {\bibinfo  {journal}
  {Phys. Rev. D}\ }\textbf {\bibinfo {volume} {108}},\ \bibinfo {pages}
  {016015} (\bibinfo {year} {2023})},\ \Eprint
  {http://arxiv.org/abs/2207.14171} {arXiv:2207.14171 [hep-ph]} \BibitemShut
  {NoStop}%
\bibitem [{\citenamefont {Sch\"afer}\ and\ \citenamefont
  {Shuryak}(1996)}]{Schafer:1995pz}%
  \BibitemOpen
  \bibfield  {author} {\bibinfo {author} {\bibfnamefont {Thomas}\ \bibnamefont
  {Sch\"afer}}\ and\ \bibinfo {author} {\bibfnamefont {Edward~V.}\ \bibnamefont
  {Shuryak}},\ }\bibfield  {title} {\enquote {\bibinfo {title} {{The
  Interacting instanton liquid in QCD at zero and finite temperature}},}\
  }\href {\doibase 10.1103/PhysRevD.53.6522} {\bibfield  {journal} {\bibinfo
  {journal} {Phys. Rev. D}\ }\textbf {\bibinfo {volume} {53}},\ \bibinfo
  {pages} {6522--6542} (\bibinfo {year} {1996})},\ \Eprint
  {http://arxiv.org/abs/hep-ph/9509337} {arXiv:hep-ph/9509337} \BibitemShut
  {NoStop}%
\bibitem [{\citenamefont {Faccioli}\ and\ \citenamefont
  {Shuryak}(2001)}]{Faccioli:2001ug}%
  \BibitemOpen
  \bibfield  {author} {\bibinfo {author} {\bibfnamefont {P.}~\bibnamefont
  {Faccioli}}\ and\ \bibinfo {author} {\bibfnamefont {Edward~V.}\ \bibnamefont
  {Shuryak}},\ }\bibfield  {title} {\enquote {\bibinfo {title} {{Systematic
  study of the single instanton approximation in QCD}},}\ }\href {\doibase
  10.1103/PhysRevD.64.114020} {\bibfield  {journal} {\bibinfo  {journal} {Phys.
  Rev. D}\ }\textbf {\bibinfo {volume} {64}},\ \bibinfo {pages} {114020}
  (\bibinfo {year} {2001})},\ \Eprint {http://arxiv.org/abs/hep-ph/0106019}
  {arXiv:hep-ph/0106019} \BibitemShut {NoStop}%
\bibitem [{\citenamefont {Shuryak}\ and\ \citenamefont
  {Zahed}(2023)}]{Shuryak:2021fsu}%
  \BibitemOpen
  \bibfield  {author} {\bibinfo {author} {\bibfnamefont {Edward}\ \bibnamefont
  {Shuryak}}\ and\ \bibinfo {author} {\bibfnamefont {Ismail}\ \bibnamefont
  {Zahed}},\ }\bibfield  {title} {\enquote {\bibinfo {title} {{Hadronic
  structure on the light front. I. Instanton effects and quark-antiquark
  effective potentials}},}\ }\href {\doibase 10.1103/PhysRevD.107.034023}
  {\bibfield  {journal} {\bibinfo  {journal} {Phys. Rev. D}\ }\textbf {\bibinfo
  {volume} {107}},\ \bibinfo {pages} {034023} (\bibinfo {year} {2023})},\
  \Eprint {http://arxiv.org/abs/2110.15927} {arXiv:2110.15927 [hep-ph]}
  \BibitemShut {NoStop}%
\bibitem [{\citenamefont {Liu}\ \emph {et~al.}(2023{\natexlab{a}})\citenamefont
  {Liu}, \citenamefont {Shuryak},\ and\ \citenamefont {Zahed}}]{Liu:2023yuj}%
  \BibitemOpen
  \bibfield  {author} {\bibinfo {author} {\bibfnamefont {Wei-Yang}\
  \bibnamefont {Liu}}, \bibinfo {author} {\bibfnamefont {Edward}\ \bibnamefont
  {Shuryak}}, \ and\ \bibinfo {author} {\bibfnamefont {Ismail}\ \bibnamefont
  {Zahed}},\ }\bibfield  {title} {\enquote {\bibinfo {title} {{Hadronic
  structure on the light-front. VII. Pions and kaons and their partonic
  distributions}},}\ }\href {\doibase 10.1103/PhysRevD.107.094024} {\bibfield
  {journal} {\bibinfo  {journal} {Phys. Rev. D}\ }\textbf {\bibinfo {volume}
  {107}},\ \bibinfo {pages} {094024} (\bibinfo {year} {2023}{\natexlab{a}})},\
  \Eprint {http://arxiv.org/abs/2302.03759} {arXiv:2302.03759 [hep-ph]}
  \BibitemShut {NoStop}%
\bibitem [{\citenamefont {Liu}\ \emph {et~al.}(2023{\natexlab{b}})\citenamefont
  {Liu}, \citenamefont {Shuryak},\ and\ \citenamefont {Zahed}}]{Liu:2023fpj}%
  \BibitemOpen
  \bibfield  {author} {\bibinfo {author} {\bibfnamefont {Wei-Yang}\
  \bibnamefont {Liu}}, \bibinfo {author} {\bibfnamefont {Edward}\ \bibnamefont
  {Shuryak}}, \ and\ \bibinfo {author} {\bibfnamefont {Ismail}\ \bibnamefont
  {Zahed}},\ }\bibfield  {title} {\enquote {\bibinfo {title} {{Hadronic
  structure on the light-front VIII. Light scalar and vector mesons}},}\
  }\href@noop {} {\  (\bibinfo {year} {2023}{\natexlab{b}})},\ \Eprint
  {http://arxiv.org/abs/2307.16302} {arXiv:2307.16302 [hep-ph]} \BibitemShut
  {NoStop}%
\bibitem [{\citenamefont {Ioffe}(2003)}]{Ioffe:2002ee}%
  \BibitemOpen
  \bibfield  {author} {\bibinfo {author} {\bibfnamefont {B.~L.}\ \bibnamefont
  {Ioffe}},\ }\bibfield  {title} {\enquote {\bibinfo {title} {{Condensates in
  quantum chromodynamics}},}\ }\href {\doibase 10.1134/1.1540654} {\bibfield
  {journal} {\bibinfo  {journal} {Phys. Atom. Nucl.}\ }\textbf {\bibinfo
  {volume} {66}},\ \bibinfo {pages} {30--43} (\bibinfo {year} {2003})},\
  \Eprint {http://arxiv.org/abs/hep-ph/0207191} {arXiv:hep-ph/0207191}
  \BibitemShut {NoStop}%
\bibitem [{\citenamefont {Diakonov}\ \emph {et~al.}(1996)\citenamefont
  {Diakonov}, \citenamefont {Polyakov},\ and\ \citenamefont
  {Weiss}}]{Diakonov:1995qy}%
  \BibitemOpen
  \bibfield  {author} {\bibinfo {author} {\bibfnamefont {Dmitri}\ \bibnamefont
  {Diakonov}}, \bibinfo {author} {\bibfnamefont {Maxim~V.}\ \bibnamefont
  {Polyakov}}, \ and\ \bibinfo {author} {\bibfnamefont {C.}~\bibnamefont
  {Weiss}},\ }\bibfield  {title} {\enquote {\bibinfo {title} {{Hadronic matrix
  elements of gluon operators in the instanton vacuum}},}\ }\href {\doibase
  10.1016/0550-3213(95)00675-3} {\bibfield  {journal} {\bibinfo  {journal}
  {Nucl. Phys. B}\ }\textbf {\bibinfo {volume} {461}},\ \bibinfo {pages}
  {539--580} (\bibinfo {year} {1996})},\ \Eprint
  {http://arxiv.org/abs/hep-ph/9510232} {arXiv:hep-ph/9510232} \BibitemShut
  {NoStop}%
\bibitem [{\citenamefont {Weiss}(2021)}]{Weiss:2021kpt}%
  \BibitemOpen
  \bibfield  {author} {\bibinfo {author} {\bibfnamefont {C.}~\bibnamefont
  {Weiss}},\ }\bibfield  {title} {\enquote {\bibinfo {title} {{Nucleon matrix
  element of Weinberg's CP-odd gluon operator from the instanton vacuum}},}\
  }\href {\doibase 10.1016/j.physletb.2021.136447} {\bibfield  {journal}
  {\bibinfo  {journal} {Phys. Lett. B}\ }\textbf {\bibinfo {volume} {819}},\
  \bibinfo {pages} {136447} (\bibinfo {year} {2021})},\ \Eprint
  {http://arxiv.org/abs/2103.13471} {arXiv:2103.13471 [hep-ph]} \BibitemShut
  {NoStop}%
\bibitem [{\citenamefont {Bodwin}\ \emph {et~al.}(2006)\citenamefont {Bodwin},
  \citenamefont {Braaten}, \citenamefont {Lee},\ and\ \citenamefont
  {Yu}}]{Bodwin:2006yd}%
  \BibitemOpen
  \bibfield  {author} {\bibinfo {author} {\bibfnamefont {Geoffrey~T.}\
  \bibnamefont {Bodwin}}, \bibinfo {author} {\bibfnamefont {Eric}\ \bibnamefont
  {Braaten}}, \bibinfo {author} {\bibfnamefont {Jungil}\ \bibnamefont {Lee}}, \
  and\ \bibinfo {author} {\bibfnamefont {Chaehyun}\ \bibnamefont {Yu}},\
  }\bibfield  {title} {\enquote {\bibinfo {title} {{Exclusive two-vector-meson
  production from e+ e- annihilation}},}\ }\href {\doibase
  10.1103/PhysRevD.74.074014} {\bibfield  {journal} {\bibinfo  {journal} {Phys.
  Rev. D}\ }\textbf {\bibinfo {volume} {74}},\ \bibinfo {pages} {074014}
  (\bibinfo {year} {2006})},\ \Eprint {http://arxiv.org/abs/hep-ph/0608200}
  {arXiv:hep-ph/0608200} \BibitemShut {NoStop}%
\bibitem [{\citenamefont {Lansberg}\ and\ \citenamefont
  {Pham}(2008)}]{Lansberg:2008cq}%
  \BibitemOpen
  \bibfield  {author} {\bibinfo {author} {\bibfnamefont {J.~P.}\ \bibnamefont
  {Lansberg}}\ and\ \bibinfo {author} {\bibfnamefont {T.~N.}\ \bibnamefont
  {Pham}},\ }\bibfield  {title} {\enquote {\bibinfo {title} {{Two-photon decay
  of pseudoscalar quarkonia}},}\ }\href {\doibase 10.1063/1.2987179} {\bibfield
   {journal} {\bibinfo  {journal} {AIP Conf. Proc.}\ }\textbf {\bibinfo
  {volume} {1038}},\ \bibinfo {pages} {259} (\bibinfo {year} {2008})},\ \Eprint
  {http://arxiv.org/abs/0804.2180} {arXiv:0804.2180 [hep-ph]} \BibitemShut
  {NoStop}%
\bibitem [{\citenamefont {Fabiano}\ and\ \citenamefont
  {Pancheri}(2003)}]{Fabiano:2002se}%
  \BibitemOpen
  \bibfield  {author} {\bibinfo {author} {\bibfnamefont {Nicola}\ \bibnamefont
  {Fabiano}}\ and\ \bibinfo {author} {\bibfnamefont {Giulia}\ \bibnamefont
  {Pancheri}},\ }\bibfield  {title} {\enquote {\bibinfo {title} {{Two photons
  width of heavy pseudoscalar mesons}},}\ }\href@noop {} {\bibfield  {journal}
  {\bibinfo  {journal} {Frascati Phys. Ser.}\ }\textbf {\bibinfo {volume}
  {31}},\ \bibinfo {pages} {417--419} (\bibinfo {year} {2003})},\ \Eprint
  {http://arxiv.org/abs/hep-ph/0210279} {arXiv:hep-ph/0210279} \BibitemShut
  {NoStop}%
\bibitem [{\citenamefont {Erler}(1999)}]{Erler:1998sy}%
  \BibitemOpen
  \bibfield  {author} {\bibinfo {author} {\bibfnamefont {Jens}\ \bibnamefont
  {Erler}},\ }\bibfield  {title} {\enquote {\bibinfo {title} {{Calculation of
  the QED coupling alpha (M(Z)) in the modified minimal subtraction scheme}},}\
  }\href {\doibase 10.1103/PhysRevD.59.054008} {\bibfield  {journal} {\bibinfo
  {journal} {Phys. Rev. D}\ }\textbf {\bibinfo {volume} {59}},\ \bibinfo
  {pages} {054008} (\bibinfo {year} {1999})},\ \Eprint
  {http://arxiv.org/abs/hep-ph/9803453} {arXiv:hep-ph/9803453} \BibitemShut
  {NoStop}%
\end{thebibliography}%

\end{document}